\documentclass[10pt]{article}

\newcommand{\f}{\begin{equation}}
\newcommand{\ff}{\end{equation}}
\newcommand{\p}{\begin{figure}[h]\begin{center}}
\newcommand{\pp}{\end{center}\end{figure}}
\usepackage[dvips]{graphicx}

\title{Supersymmetric Spin Networks and Quantum Supergravity}

\author{Yi Ling$^{\dagger}$ and Lee Smolin*
\thanks{*smolin@phys.psu.edu, \dag ling@phys.psu.edu}\\
\it *Center for Gravitational Physics and Geometry,Department of
Physics\\ \it Pennsylvania State University, University Park , PA
16802\\ \it ${}^{\dagger}$Department of Physics, Pennsylvania State
University\\ \it University Park, PA 16802\normalsize}

\begin{document}

\maketitle

\begin{abstract}

We define supersymmetric spin networks, which provide a complete
set of gauge invariant states for supergravity and supersymmetric
gauge theories. The particular case of $Osp(1|2)$ is studied in
detail and applied to the non-perturbative quantization of
supergravity.  The supersymmetric extension of the area operator
is defined and partly diagonalized.  The spectrum is discrete
as in quantum general relativity, and the two cases could
be distinguished by measurements of quantum geometry.

\end{abstract}

\section{Introduction}

\noindent In this paper we describe an extension of the spin
network states to supergravity. The spin network states play a
fundamental role in non-perturbative quantizations of both gauge
theories \cite{KS,bae1} and gravitational theories
\cite{spain,sn1,sn2}. In the gauge theory context they provide an
orthonormal basis for lattice gauge theories \cite{KS,bae1}. In this
case the spin networks are labeled graphs on the lattice, whose
edges are labeled by the finite irreducible representations of the
gauge group $G$.  In quantum gravity diffeomorphism invariance
reduces the degrees of freedom, so that a basis of states
invariant under spatial diffeomorphisms and local frame rotations
are given by the diffeomorphism classes of spin networks
\cite{sn1,sn2}. In this case the group is $SU(2)$, for the chiral
formulation based on the Ashtekar-Sen variables \cite{sen,abhay},
or $SU(2)+SU(2)$ in the relativistic
case\cite{barrettcrane1,hologr}.

Over the last ten years there  has been a great deal of progress
in our understanding of the non-perturbative structure of quantum
general relativity, leading to the complete formulation of the
quantum theory\footnote{For recent reviews see
\cite{carlo-review,future}.}. Among the key results
are the discovery that diffeomorphism invariant observables that
measure aspects of the spatial geometry such as areas of surfaces
and volumes of regions are finite, and have discrete, computatable
spectra\cite{spain,sn1,sn2,vol2}.
This has led to a physical understanding of the spin
network states as eigenstates of these geometrical
observables.

During this period there have been a number of papers which extend
the methods used to
supergravity\cite{superstuff,GSU1,GSU2,SUG2,N=4}.  These have
included the formulation of ${\cal N}=1,2$\cite{GSU1,GSU2,SUG2},
and ${\cal
N}=4$\cite{N=4}  supergravity in terms of chiral, Ashtekar-Sen
like variables, as well as the discovery of exact solutions to the
quantum constraints\cite{GSU1,SUG2}. However, much more remains to
be done in this direction.  The non-perturbative quantization of
gravitational theories with extended supersymmetry is largely
unexplored territory, despite the fact that extended supersymmetry
is essential to the success of string theory, which remains the
only successful technique for investigating quantum gravity in the
perturbative regime.  Another important open area of investigation
is the properties of $BPS$ states in the non-perturbative regime.
This could be very interesting as it could provide a way to
compare results on black hole entropy gotten by both string
theory\cite{string-bh} and loop quantum gravity\cite{kirill-bh}.

In this paper we take a first step to the study of the
non-perturbative quantization of supersymmetric theories of
gravitation by constructing the spin network states for $N=1$
supergravity.  We find a number of new features, which suggest
that this could be a fruitful direction of investigation.  The
main result is a diagrammatic method for the construction and
evaluation of spin networks for the supergroup $Osp(1|2)$.  As a
first example we construct and partly diagonalize the
supersymmetric extension of the area operator.  As expected the
spectrum is discrete, but different from that of quantum general
relativity. This means that experimental probes of geometry at the
Planck scale could in principle distinguish different hypotheses
about the local gauge symmetry. This is highly interesting in
light of recent developments that suggest that astrophysical
probes of Planck scale physics can be developed\cite{giovanni}.

Another possible application of the formalism given here is to
supersymmetric Yang-Mills theory.  It will be very interesting to
investigate the extent to which the physics of $N=2$ and $N=4$
super-Yang-Mills theory can be expressed in terms of the spin network
states.

It is straightforward to extend the construction here to $N=2$ and
higher supersymmetry, this will be described elsewhere \cite{yi2,
yi3}. Also, in progress \cite{SUG1} is an examination of the
canonical and boundary structure of $N=1,2$ quantum supergravity,
which extends results on a holographic formulation of quantum
general relativity with finite cosmological constant
\cite{linking,hologr}.

In the next section we review some of the basic results about
$N=1$ supergravity in chiral coordinates, first studied by
Jacobson\cite{superstuff}. In section 3 we present some results
from the representation theory of $Osp(1|2)$ which allow us in
section 4 to construct quantum spin networks.  The diagrammatic
method for doing calculations with these states is introduced in
section 5, and the following sections describe examples and
calculations.

Finally, we mention that we do not here provide rigorous proofs for
the assertions made, but we see no reason why a straightforward
extension of the rigorous methods
introduced in \cite{rayner,chrisabhay,gangof5,thomas} to the present
case should not be possible.

\section{Review of Quantum Supergravity}

Supergravity in terms of the new variables maybe was initially
investigated in \cite{superstuff} and extended in \cite{GSU1,GSU2}.
 In this paper we will consider mainly $N=1$ supergravity. As
shown first by Jacobson in \cite{superstuff}, this can be
formulated in chiral variables which extend the Ashtekar-Sen
variables of general relativity.  In this formulation, the
canonical variables are the left handed $su(2)$ spin connection
$A_a^i$ and its superpartner spin-3/2 field $\psi_a^A$. As shown
in \cite{SUG2} these fit together into a connection field of the
superlie algebra $Osp(1|2)$ (which is referred to in some
references \cite{GSU1,SUG2,GSU3} as $GSU(2)$.)

We thus define the graded connection:
\f
\cal{A}\mit_a:=A_a^iJ_i+\psi_a^AQ_A
\ff
where $a$ is the spatial index. If $\widetilde{E^a_i}$ and
$\pi^a_A$ are momenta of $A_a^i$ and $\psi_a^A$ respectively, we
can define the graded momentum as:
\f
\cal{E}\mit^a:=\widetilde{E^a_i}J^i+\pi^a_AQ^A
\ff
The constraints that generate local gauge
transformations can then be expressed as usual as,
\f
{\cal G}_{i} =
D_a{\widetilde{E^a_i}}+\frac{i}{\sqrt{2}}\pi^a_A\psi_{aB}\tau^{AB}_i=0
\label{gauss}
\ff
The left and right handed supersymmetry transformations are
generated by\cite{superstuff},
\f
{\cal L}_{A} =
D_a\pi^a_A-ig{\widetilde{E^a_i}}\tau^B_{iA}\psi_{aB}=0 \label{lss}
\ff
\f
{\cal R}^{A} =
\epsilon^{ijk}\widetilde{E^a_i}\widetilde{E^b_j}\sigma^A_{kB}
(-4iD_{[a}\psi^B_{b]} +\sqrt{2} g\epsilon_{abc}\pi^{cB})=0
\label{rss}
\ff
where the cosmological constant is given by $\Lambda=-g^2$. The
diffeomorphism and hamiltonian constraints can be derived by
taking the Poisson Brackets of (4) and (5).

These may be written simply in terms of the fundamental
representation of $Osp(1|2)$, which is $3$ dimensional.  The
superlie algebra $Osp(1|2)$ is then generated by five $3\times 3$
matrices $G_I(I=1...5)$, given explicitly in \cite{SUG2}. Using
them we can define \f \cal{A}\mit_a^I=(A_a^i,\psi_a^A) \ff \f
\cal{E}\mit^a_I=(\widetilde{E^a_i},\pi^a_A) \ff where $I=(i,A)$
labels the five generators of $Osp(1|2)$.

Then the first two constraints can be combined into one $Osp(1|2)$
Gauss constraint: \f \cal{D}\mit_a\cal{E}\mit^a_I=0 \label{sgauss}
\ff
while the last one combines with the
Hamiltonian constraint to give:

\f \cal{E}\mit^a\cal{E}\mit^b\cal{F}\mit_{ab}-ig^2
\epsilon_{abc}\cal{E}\mit^a\cal{E}\mit^b \cal{E}\mit^c=0 \ff where
$\cal{F}\mit_{ab}$ is the curvature of the super connection
$\cal{A}\mit_a$ : \f
\cal{F}\mit_{ab}:=d_a\cal{A}\mit_b+[\cal{A}\mit_a,\cal{A}\mit_b]
\ff

A key feature of this kind of approach to supergravity is that the
supersymmetry gauge invariance has been split into two parts,
which play rather different roles.  The left handed supersymmetry
transformations generated by eq. (\ref{lss}) combine with the
$SU(2)$ Gauss's law (\ref{gauss}) to give a local $Osp(1|2)_{L}$
gauge invariance.  The theory is then written so that the
associated $Osp(1|2)$ connection is the canonical coordinate.  The
right handed part of the supersymmetry, generated by (\ref{rss}),
is a dynamical constraint,  being quadratic rather than linear in
the momentum. It  joins with the Hamiltonian constraint
to form a left handed supersymmetry multiplet of
dynamical operators.

It is then natural in a chiral formulation of supergravity to
represent the left-handed supersymmetry kinematically, and solve
it completely by expressing the theory completely in terms of
$Osp(1|2)$ invariant states.  These are the spin network states we
will present shortly.  The remaining, right handed, part of the
supersymmetry is then imposed as a dynamical operator, and has the
same status as the Hamiltonian constraint.

The loop representation for supergravity in the chiral
representation was constructed in \cite{SUG2} in terms of
$Osp(1|2)$ Wilson loops.  These are defined in terms of the
supertrace taken in the fundamental $3$ dimensional representation
of $Osp(1|2)$.
\f
\cal T\mit[\gamma]=Str\cal P\mit
exp(\oint_{\gamma}ds \cal A\mit_a {\gamma}^a) \equiv Str{ \cal
U\mit}_{\gamma}({\cal A\mit })
\ff

These Wilson loop states are subject to additional relations
arising from intersections of loops.  These are solved completely
by the introduction of the spin network basis, which are complete
and orthogonal\cite{sn2}.

We can construct the loop-momentum variables by inserting the
$Osp(1|2)$ invariant momentum $\cal E\mit^a$ into the Wilson loops: \f
\cal T\mit^a[\alpha](s)=Str[\cal U\mit_{\alpha}(\cal A\mit)\cal
E\mit^a(\alpha(s)] \ff
It is straightforward to show that the
$\cal T\mit[\gamma]$ and $\cal T\mit^a[\alpha](s)$ form a closed
algebra under Poisson brackets, which we will call the $N=1$
super-loop algebra.

We will also need to describe operators quadratic in the conjugate
momenta, which in the loop representation are formed by inserting
two momenta in the loop trace,
\f
\cal T\mit^{ab}[\alpha](s,t)=Str[\cal U\mit_{\alpha}(s,t)\cal
E\mit^a(\alpha(t)) \cal U\mit_{\alpha}(t,s)\cal
E\mit^b(\alpha(s))]
\ff
The higher order loop operators are similarly defined as
\f
\cal T\mit^{ab...c}[\alpha](s,t,...v)=Str[\cal U\mit_{\alpha}(s,t)\cal
E\mit^a(\alpha(t)) \cal U\mit_{\alpha}(t,u)\cal
E\mit^b(\alpha(u))...\cal U\mit_{\alpha}(v,s) \cal
E\mit^c(\alpha(s))]
\ff

As discussed in \cite{SUG2}, the supersymmetric extension of the
Chern-Simons state may be formed from the Chern-Simons form of the
superconnection ${\cal A}_{a}$,

\f
\Psi_{SCS}(\cal{A}\mit_a)=exp[\frac{i}{2\Lambda}\int
d^3xSTr(\cal{A}\mit\wedge\cal{F}\mit-\frac{1}{3}
\cal{A}\mit\wedge\cal{A}\mit\wedge\cal{A}\mit)]
\ff

This state is an exact solution to all the quantum constraints.
Like the ordinary Chern-Simons state it also has a semiclassical
interpretation as the ground state associated with DeSitter or
Anti-DeSitter spacetime.

\section{Finite Dimensional Irreducible Representation of $Osp(1|2)$.}

Spin networks may be constructed for any Lie or Superlie algebra,
$\cal A$ by extending the original definition \cite{pen1,bae1}. An
$\cal A$-spin network is a labeled graph whose edges are labeled
by the finite dimensional irreducible representations of $\cal A$
and whose nodes are labeled by the associated intertwiners.  In
quantum gravity spin networks states are associated with the gauge
group of the connection, which we have seen in the case of $N=1$
supergravity in the chiral representation \cite{superstuff} to be
$Osp(1|2)$. The representation theory of $Osp(1|2)$ has been
studied in detail in \cite{GSU3,GSU4,GSU5,GSU7}, here we give some
of the basic facts that we will need to construct the associated
spin network states.

The superlie algebra of $Osp(1|2)$,  is constructed by three
bosonic generators $J_i$ (i=1,2,3)and two fermionic generators
$Q_A(A=0,1)$. The commutation relations are :

\f
[J_i,J_j]=i\epsilon_{ijk}J_k
\ff
\f
[J_i,Q_A]=1/2(\tau_i)_A^BQ_B
\ff
\f
\{Q_A,Q_B\}=1/2\epsilon_{AB}\tau^iJ_i
\ff
where $\tau^i$ are Pauli matrices.

Each irreducible representation of the $Osp(1|2)$ contains two
adjacent $SU(2)$ representations. One is labeled by spin $J$ and
the other by $J-1/2$.  $J$ may be taken as the label of the
$Osp(1|2)$ representation, and is related to the eigenvalues of
the quadratic Casmier operator of the supergroup $Osp(1|2)$: \f
C=J^iJ_i+\epsilon^{AB}Q_AQ_B \ff by \f \hat{C}|J>=J(J+1/2)|J> \ff

For each $J$ the representation is a graded vector space with a basis
labeled by $|J;L;M >$,
where J is an integer or half-integer, and L can be
$J$ or $J-1/2$ and $-L\le M\le L$.
The dimension of the space of the representation with spin
$J$ is $4J+1$.

The usual rules for combination of angular momentum can be
extended directly to these states.  The result is a super
Racah-Wigner calculus which gives the results of decompositions of
products of representations of $Osp(1|2)$.  The tensor product is
completely reducible and is given by \f
j_1\textstyle{\bigotimes} j_2=|j_1-j_2|\textstyle{\bigoplus}
|j_1-j_2+1/2|\textstyle{\bigoplus} ...|j_1+j_2| \label{tp} \ff

Note that this
differs from the familiar $SU(2)$ case in that representations
which differ from
$j_{1}+j_{2}$ by both integers and half integers are included.
The Clebsch-Gordon
coefficients for the expansion of the basis elements are
determined uniquely, from
their values for $SU(2)$.

Next we consider the reduction of the tensor product of three
irreducible representations $(j_1,j_2,j_3)$.  As in the $SU(2)$
case we have two different recoupling schemes. One can couple the
representations $(j_1,j_2)$ into $j_{12}$ first and then couple
the result to $j_3$ to give the final representation; or one
couples $(j_2,j_3)$ into $j_{23}$ first and then couples to $j_1$
next. These two representations are related to each other by the
Racah sum rule in terms of super rotation 6-symbols. For
$Osp(1|2)$, the parity independent super-rotation 6-symbols are
defined as[21]:

\f
{\left\{\begin{array}{ccc}
j_1&j_2&j_{12}\\j_3&j&j_{23}\end{array}\right\}}^s=
(-1)^{\Phi(\lambda_1,\lambda_2,\lambda_3)}
{\left\{\begin{array}{ccc}
j_1\lambda_1&j_2\lambda_2&j_{12}\lambda_{12}\\j_3\lambda_3&j\lambda&j_{23}
\lambda_{23}
\end{array}\right\}}^s
\ff

Then the Racah sum rule reads:
\f
{\left\{\begin{array}{ccc}
j_1&j_2&j_{12}\\j_3&j&j_{23}\end{array}\right\}}^s=
\sum_{j_{13}}(-1)^{\Theta}
{\left\{\begin{array}{ccc}
j_1&j_3&j_{13}\\j_2&j&j_{23}\end{array}\right\}}^s
{\left\{\begin{array}{ccc}
j_2&j_1&j_{12}\\j_3&j&j_{13}\end{array}\right\}}^s
\ff

In a similar way the Biedenharn-Elliott identity can be constructed for
the super 6-j symbols:
\begin{eqnarray}
&& {\left\{\begin{array}{ccc}
j_1&j_2&j_{12}\\j_3&j_{123}&j_{23}\end{array}\right\}}^s
{\left\{\begin{array}{ccc}
j_{23}&j_1&j_{123}\\j_4&j&j_{14}\end{array}\right\}}^s\nonumber\\
&&=\sum_{j_{124}}(-1)^{\theta_{be}} {\left\{\begin{array}{ccc}
j_2&j_1&j_{12}\\j_4&j_{124}&j_{14}\end{array} \right\}}^s
{\left\{\begin{array}{ccc}
j_3&j_{12}&j_{123}\\j_4&j&j_{124}\end{array}\right\}}^s
{\left\{\begin{array}{ccc}
j_{14}&j_2&j_{124}\\j_3&j&j_{23}\end{array}\right\}}^s
\end{eqnarray}
where $\theta_{be}$ and $\Theta$ are the sign factors related to
the super-spins involved. The interesting fact is that except
these two sign factors, the structure of two relations are the
same as the structure for SU(2) rotation algebra. Therefore when
we restrict the sum of three super-spins in all the triangles
$(j_1,j_2,j_{12})$, $(j_1,j_3,j_{13})$, $(j_2,j_3,j_{23})$ to be
integers, then all the expressions go back to the normal racah sum
rule and the Biedenharn-Elliott identity for $SU(2)$.

\section {Spin Network States of $N=1$ Supergravity}

We recall that a spin network state of quantum general relativity,
denoted $|\Gamma >$ consists of an embedding of closed graph
$\Gamma$ into a fixed three manifold $\Sigma$ with edges labeled
by the representation of SU(2)($SU(2)_{q}$) and vertices labeled
by intertwining operators, namely distinct ways to decompose the
incoming representations into a singlet. Here we define the {\it
super} spin networks in the same way only by replacing the $SU(2)$
with superlie algebra $Osp(1|2)$.\footnote{It is also possible to
extend the construction to the ``quantum graded group,'' $Osp(1|2)_{q}$.
We do not carry this out here.} The elements
of the super spin networks are links and
vertices. Notice that in quantum general relativity, the link of
color $n$ corresponds to a parallel propogator of connection $A_a$
along this path in the spin n/2 representation of su(2), here
associated to very link we also label a color $n_i$ which is two
times as the superspin $j_i$ which labels the representation of
$Osp(1|2)$. For every vertex $v_e$, there are incoming links with
color $n^{in}_{ei}$ and outcoming links with color $n^{out}_{ei}$.
so we can label the vertex by  the total color
$v_e=\sum_{i}n^{in}_{ei}-\sum_{i}n^{out}_{ei}$ which satisfies
$1/2\le v_e\le k/2$.

Corresponding to every super spin network $(\Gamma^{sg},n_i,v_e)$,
there is a super spin network state $<\Gamma^{sg},n_i,v_e|$ in the
Hilbert space of the supergravity. As an independent basis, the
super spin network can also be expressed as the bras so that a
general state in this representation is given by: \f
\Psi[\Gamma^{sg}]:=<\Gamma^{sg}|\Psi> \ff

We now list some basic facts about the super spin networks .
\begin{itemize}
\item{}  As in the $SU(2)$ case there is no intertwiner associated to
trivalent nodes because
given
\f
|J_1-J_2|\le J_3\le |J_1+J_2|
\label{triangle}
\ff
then the map from the tensor products of two
representations to the reduced representation
is unique.

\item{} The condition that the sum of three colors of the links adjacent
to a trivalent vertex
must sum to an even number does not hold, because both even and odd
spins appear in the sum (\ref{tp}). As the color of the link is
twice the spin,
the edges of a trivalent vertex
can be any integers such that eq. (\ref{triangle}) holds.

\item{} If the valences adjacent to the same
vertex is more than three, intertwiners are needed to label the
different maps from the incoming representations to the singlet
state. As in the $SU(2)$ case the multi-valent vertices can be
decomposed in terms of trivalent vertex  connected by internal
edges, as described in \cite{sn2}.

\end{itemize}

Note that this means that there is no simple way to decompose the
$Osp(1|2)$ spin networks completely in terms of ordinary spin
networks because there is no ordinary spin network vertex
corresponding to the superspin network vertices where the sum of
incident colors is odd. However, there is still a very useful
decomposition, which we will give below.

As in the $SU(2)$ case, there is a recoupling theory based on the
Racah sum rule and
Biedenharn-elliott identity in terms of the {\it super}-rotation
6-j symbols. We can
express the recoupling theory by fig.1:
\p
\includegraphics[angle=0,width=14cm,height=4cm]{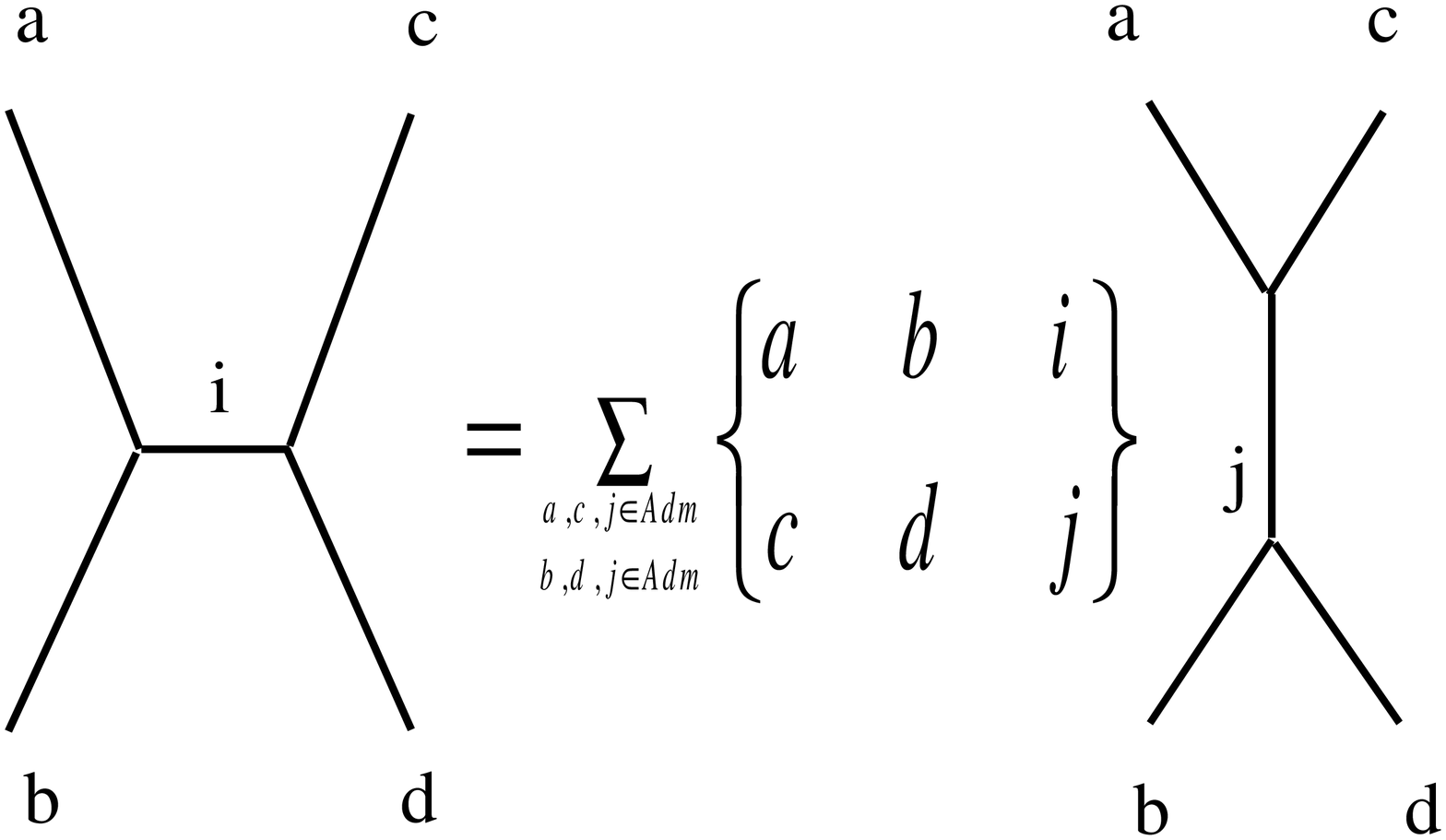}
\caption{Recoupling theory for $Osp(1|2)$ 6-j symbols} \pp

Where the sum is over labels such that the super 6-j symbols
satisfies the triangle inequalities eq. (\ref{triangle}).

\section{Graphic Representation of the Super Spin Networks}

We can now give a diagrammatic notation for $Osp(1|2)$ spin
networks which is useful for computation.  We follow the method
developed in  \cite{sn2} and elaborated in \cite{vol2} for quantum
general relativity in which a diagrammatic notion for $SU(2)$ spin
network states was developed by modifying notations used by
Penrose\cite{pen1} and Kauffman and Linns\cite{KL}. The result is
a diagrammatic notation of super spin networks based on the
connection between them and the representation theory of the
supergroup $Osp(1|2)$.

\subsection{Element of the Diagrams}

The basic fact about the $SU(2)$ representation theory on which
the Penrose and Kauffman and Linns notation rests is that all
irreducible representations can be obtained by symmetrizing
products of the fundamental representation.  In the case of
$Osp(1|2)$ all irreducible representations can be obtained via a
process of graded symmetrizing, in which there are extra signs for
even and odd parts of the representations.  There are in fact two
different fundamental representations for the $Osp(1|2)$, which
are complex conjugates of each other.  Let us consider first the
left handed fundamental representation. It is a three dimensional
graded vector space, whose elements may be written \f
\xi_\alpha=(\begin{array}{c} \psi_A \\ \phi_{\circ}
\end{array})
\ff where $A=( 0,1)$ denotes the left handed $SU(2)$  spinor
index part and $\alpha= \circ$ denotes the third component. Here
we take the $\xi_{A}= \psi_A$ to be fermionic while the
$\xi_{\circ} = \phi$ is bosonic. The grade of the index,
$g(\alpha)$, is defined to be one for $\circ$ and zero for $A$.
Under the action of $Osp(1|2)$, $\xi_\alpha$ transforms as: \f
\xi_{\alpha'}={U_{\alpha'}}^\alpha\xi_{\alpha} \ff where $3\times
3$ matrix $U_{\alpha'}^\alpha$ is an element in the fundamental
representation of $Osp(1|2)$.

The higher irreducible representations are formed by taking graded
symmetric products of this fundamental representation.
For instance the basis states for J=1 span a five dimensional space,
which can be
constructed by symmetrized tensor products of two states in the
fundamental representation, as,
\f
\xi_{(\alpha\beta)}:=\frac{1}{2}[\xi_{\alpha}\xi_{\beta}+(-1)^{g(\alpha)g(\beta)}
\xi_{\beta}\xi_{\alpha}]
\ff

We can then read off the components of the basis states of the $J=1$
representation.  They consist of a pair of $SU(2)$ representations,
given by,
\f
\xi_{(\alpha\beta)}=(\xi_{(AB)},\xi_{(A}\phi_{\circ)})
\ff
The first term is the bosonic component defined as,
\f
\xi_{(AB)}=\frac{1}{2}(\psi_A^{(1)}\psi_B^{(2)}+\psi_B^{(1)}\psi_A^{(2)})
\ff
and the second term is the fermionic component of the basis states.
\f
\xi_{(A}\phi_{\circ)}=\frac{1}{2}(\psi_A^{(1)}
\phi_\circ^{(2)}+\phi_\circ^{(1)}\psi_A^{(2)})
\ff
The other term $\phi_{[\circ \circ]}$ vanishes due to the antisymmetrization.

Under the action of $Osp(1|2)$, the states transform as: \f
\xi_{(\alpha'\beta')}={U_{(\alpha'\beta')}}^{(\alpha\beta)}\xi_{\alpha\beta}
\ff where:
\begin{eqnarray}
{U_{(\alpha'\beta')}}^{(\alpha\beta)}&=&\frac{1}{2}[(-1)^{g(\alpha)
[g(\beta')-g(\beta)]}
{U_{\alpha'}}^{\alpha}{U_{\beta'}}^{\beta}\nonumber\\&&+(-1)^{g(\alpha')
g(\beta')}(-1)^{g(\alpha)[g(\alpha')-g(\beta)]}
{U_{\beta'}}^{\alpha}{U_{\alpha'}}^{\beta}] \end{eqnarray}

If we only consider the unit element of $Osp(1|2)$ in this
representation, then we have \f
{\delta_{(\alpha'\beta')}}^{(\alpha\beta)}=\frac{1}{2}[
{\delta_{\alpha'}}^{\alpha}{\delta_{\beta'}}^{\beta}+(-1)^{g(\alpha')g(\beta')}
{\delta_{\beta'}}^{\alpha}\delta_{\alpha'}^{\beta}] \ff This
allows us to generalize the Penrose diagrammatic notation for
$SU(2)$ spin networks. We indicate the elements of  a super spin
networks by bold lines, the elements with su(2) indices by thin
lines and third component $\phi=\xi_{\circ}$ by dotted lines. Then
we can denote the $\delta_{\alpha'}^{\alpha}$ and its components
as fig.2.
\p
\includegraphics[angle=0,width=8cm,height=4cm]{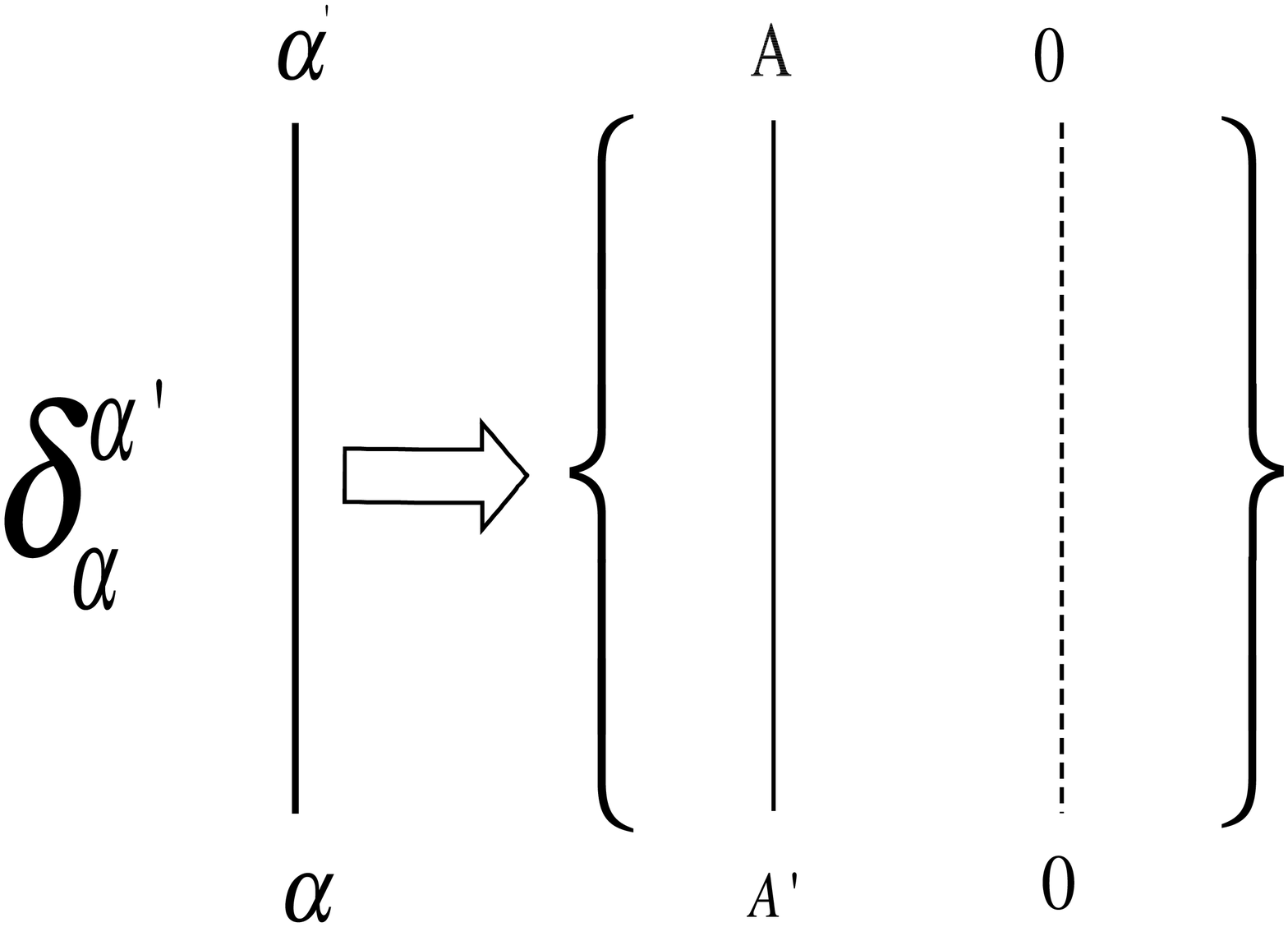}
\caption{Unit element in fundamental representation of $Osp(1|2)$}
\pp

Then it's straightforward to express (35) as fig.3.
\p
\includegraphics[angle=0,width=8cm,height=4cm]{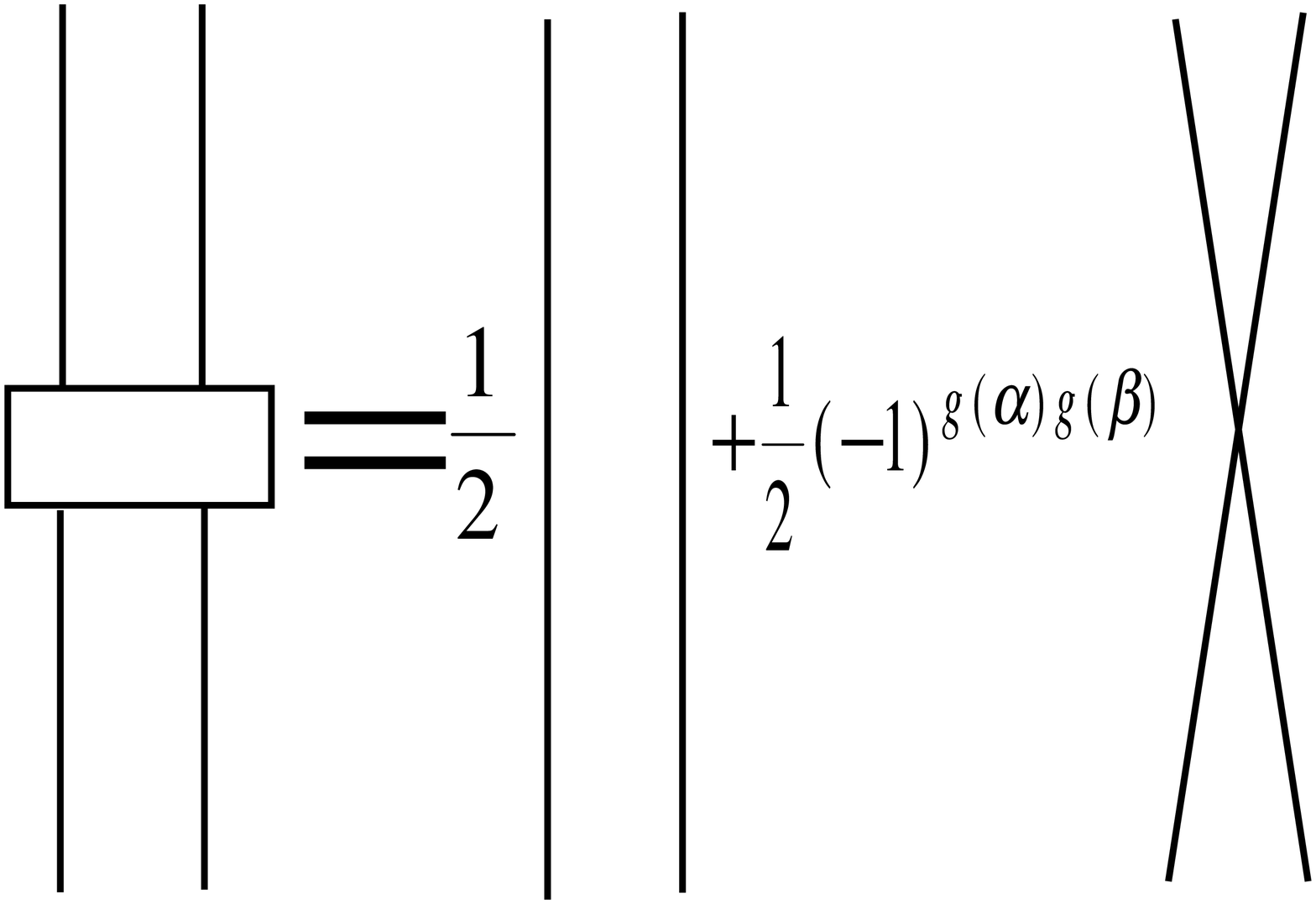}
\caption{Construction of unit element in representation with spin one}
\pp

Let us consider the component formulation of this expression. When
the indices of delta are $SU(2)$ spinor indices, it's easy to see
that it goes back to the normal spin networks expression. If
one index is fermionic and the other one is bosonic, they
commute with other and we can denote the expression by two vertical lines,
one solid and one dotted. If both indices are bosonic, which may
be denoted by two vertical dotted lines.  However this term
vanishes because the graded symmetrization antisymmetrizes them
and there is a single bosonic component. The procedure of the
decomposition of the super element can then be drawn as fig.4.
\p
\includegraphics[angle=0,width=8cm,height=4cm]{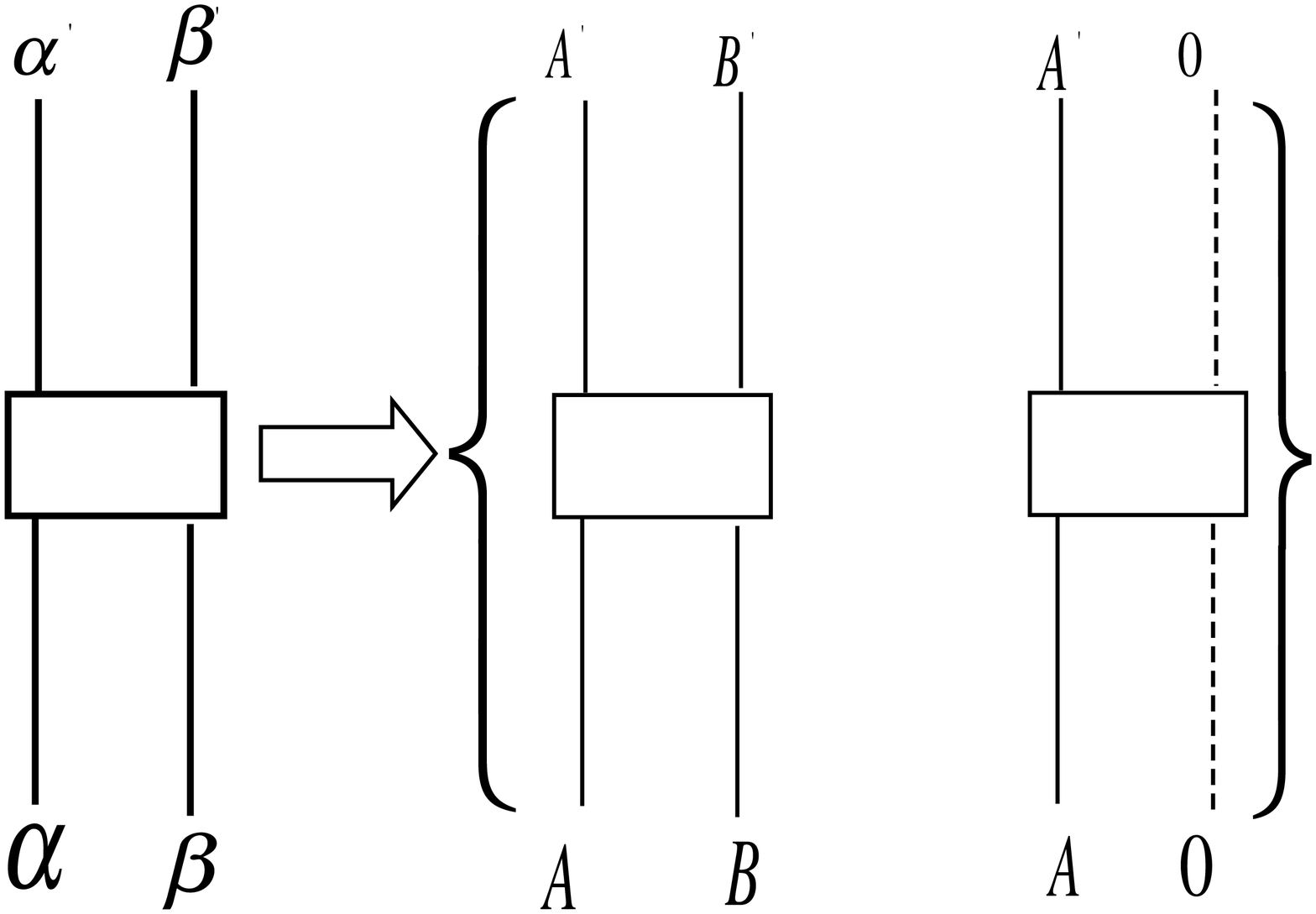}
\caption{Decomposition of the link with color 2 in super spin
networks into normal SU(2) ones} \pp

Thus, we have a way to decompose the
diagrams for $Osp(1|2)$ spin networks into combinations of $SU(2)$
spin network diagrams and dotted lines representing the single
bosonic component of the fundamental representation. It is
straightforward to see how this works when applied to any higher
dimensional representation of $Osp(1|2)$, which is gotten by
making a graded symmetrization of $n$ fundamental representations.
The basic property is that all the terms whose corresponding
graphs have two or more dotted lines must vanish also since we
need to antisymmetrize them. As a result, the basis states in any
dimensional representation consists of two components, \f
\xi_{(\alpha_1...\alpha_n)}=(\xi_{(A_1...A_n)},\xi_{(A_1...A_{n-1}0)})
, \ff where \f
\xi_{(A_1...A_n)}=\psi_{(A_1}\psi_{A_2}...\psi_{A_n)} \ff \f
\xi_{(A_1...A_{n-1}0)}=\psi_{(A_1}\psi_{A_2}...\psi_{A_n-1}\phi_{0)}
. \ff

The unit element of the supergroup in this representation can be
expressed as:
\f
{\delta_{(\alpha'\beta'...\gamma')}}^{(\alpha\beta...\gamma)}:=
{\delta_{(\alpha'}}^{\alpha}
{\delta_{\beta'}}^{\beta}...{\delta_{\gamma')}}^{\gamma}
\ff

and the corresponding graph can be drawn as fig.5. Thus we see
that we can decompose a super spin network into a sum of diagrams,
each of which is a normal spin network together with dotted lines.
In this decomposition each edge of the superspin network, with
color $n$ becomes two ordinary spin network edges, the first an
$n$ line without a dotted line and the second with an $n-1$ line
with a single dotted line. This is shown in fig.5.
\p
\includegraphics[angle=0,width=8cm,height=4cm]{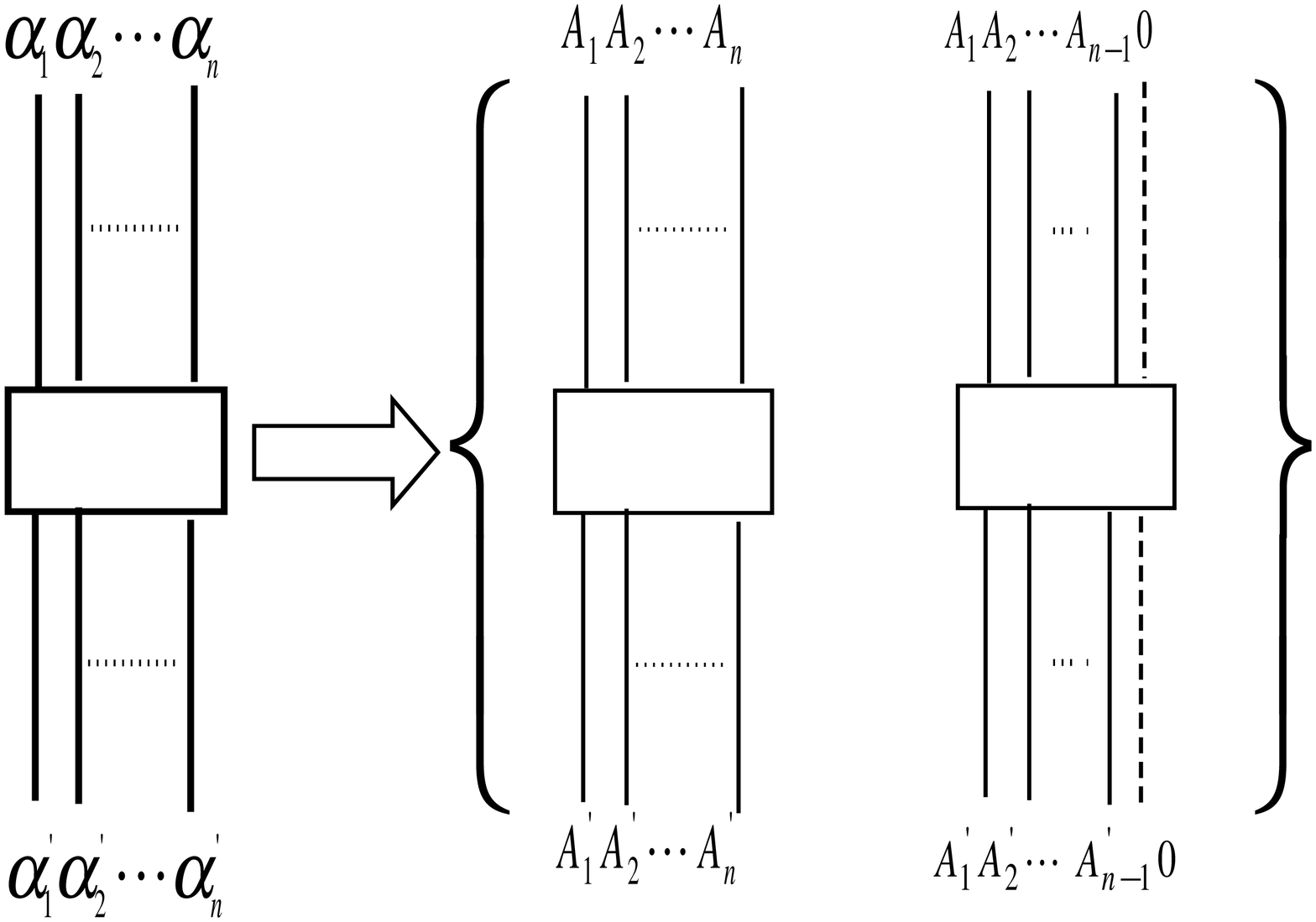}
\caption{Decomposition of the super link of color $n$}
\pp

\subsection{Trivalent Vertices}

Next we consider the tri-valent vertex. As there is no restriction
that the incident colors must add up to an even number, as in the
$SU(2)$ case, the simplest trivalent node is the one in which all
three edges have color one. This node can be visualized in two
ways, depending on how the direction of time is read. One fermion
with spin one half meets one boson with spin zero and then changes
into one fermion, or two fermions with spin one half meet together
forming into a boson which is also singlet state. These processes
are expressed by fig.6.

\p
\includegraphics[angle=0,width=8cm,height=4cm]{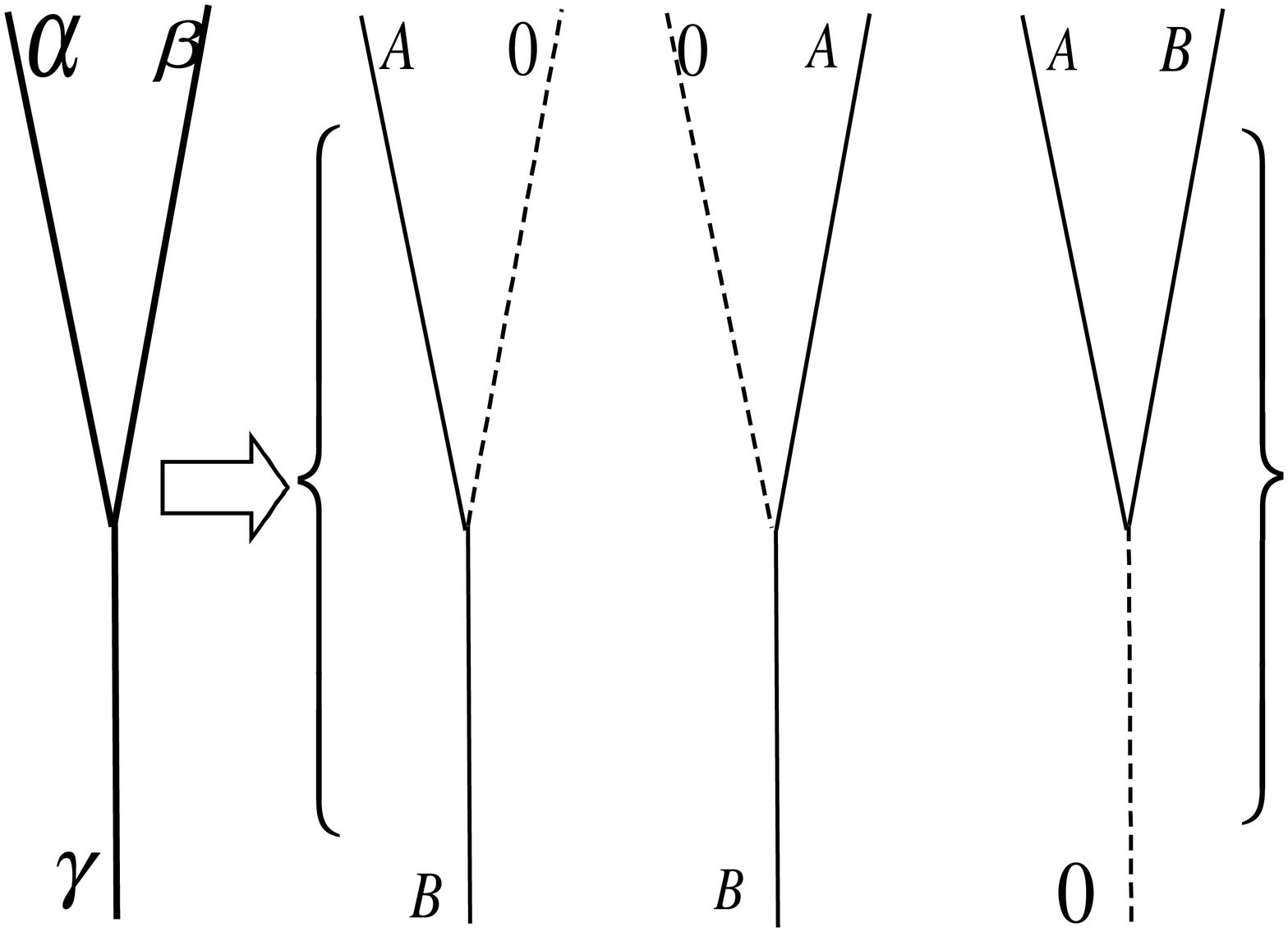}
\caption{The three terms in the decomposition of the trivalent
vertex, in the case that all colors are equal to one.}
\pp

We next consider the case in which every link has color two. This
can be decomposed into the ordinary $SU(2)$ spin networks as shown
in fig.7.
\p
\includegraphics[angle=0,width=8cm,height=5cm]{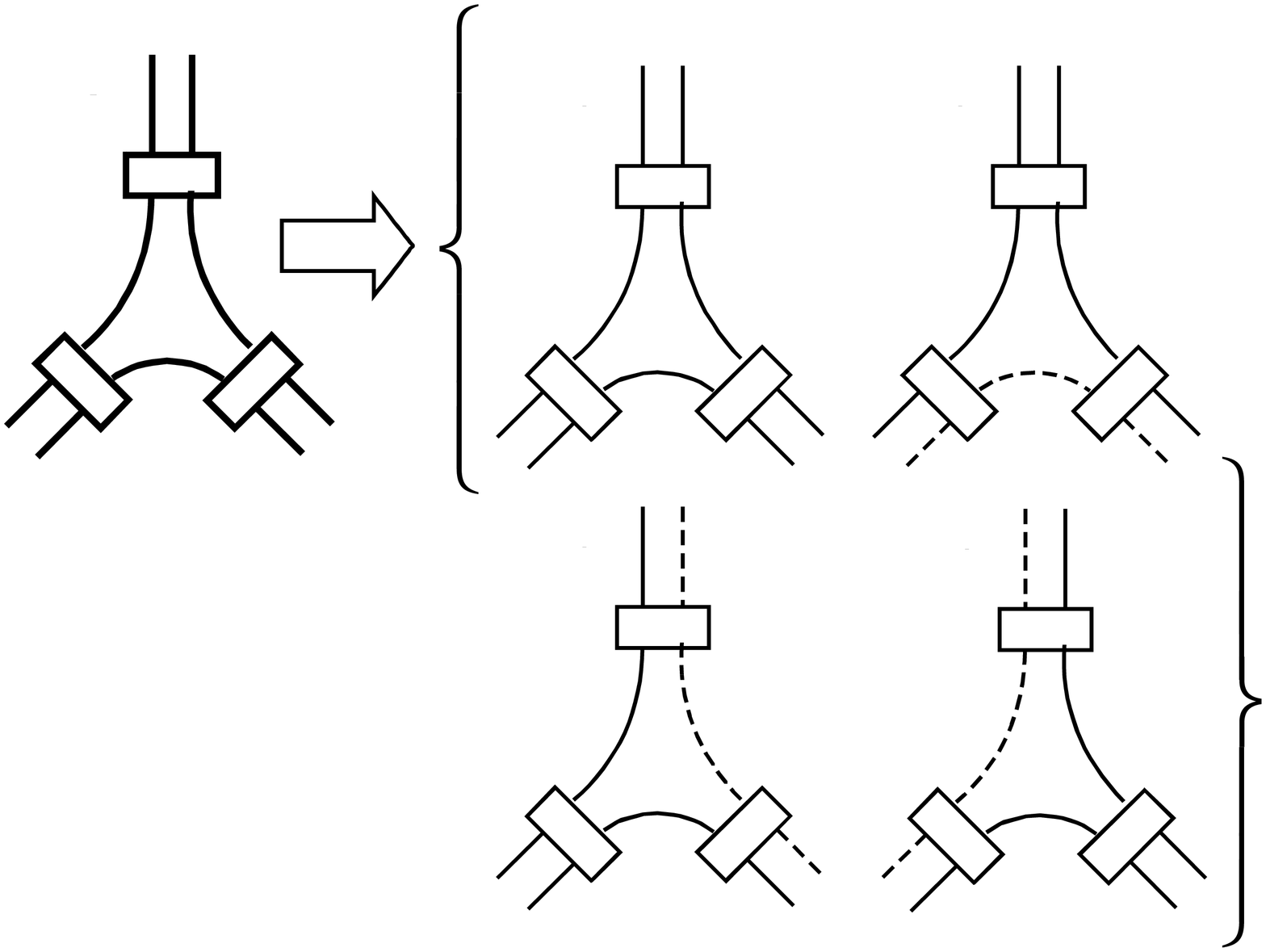}
\caption{ Decomposition of trivalent vertex in which every link
has color two.}
\pp

In general, if the sum of the three colors is even it can be
decomposed into four terms, each of which contains an ordinary
spin network plus, possible dotted edges. We illustrate it in
fig.8.
\p
\includegraphics[angle=0,width=8cm,height=5cm]{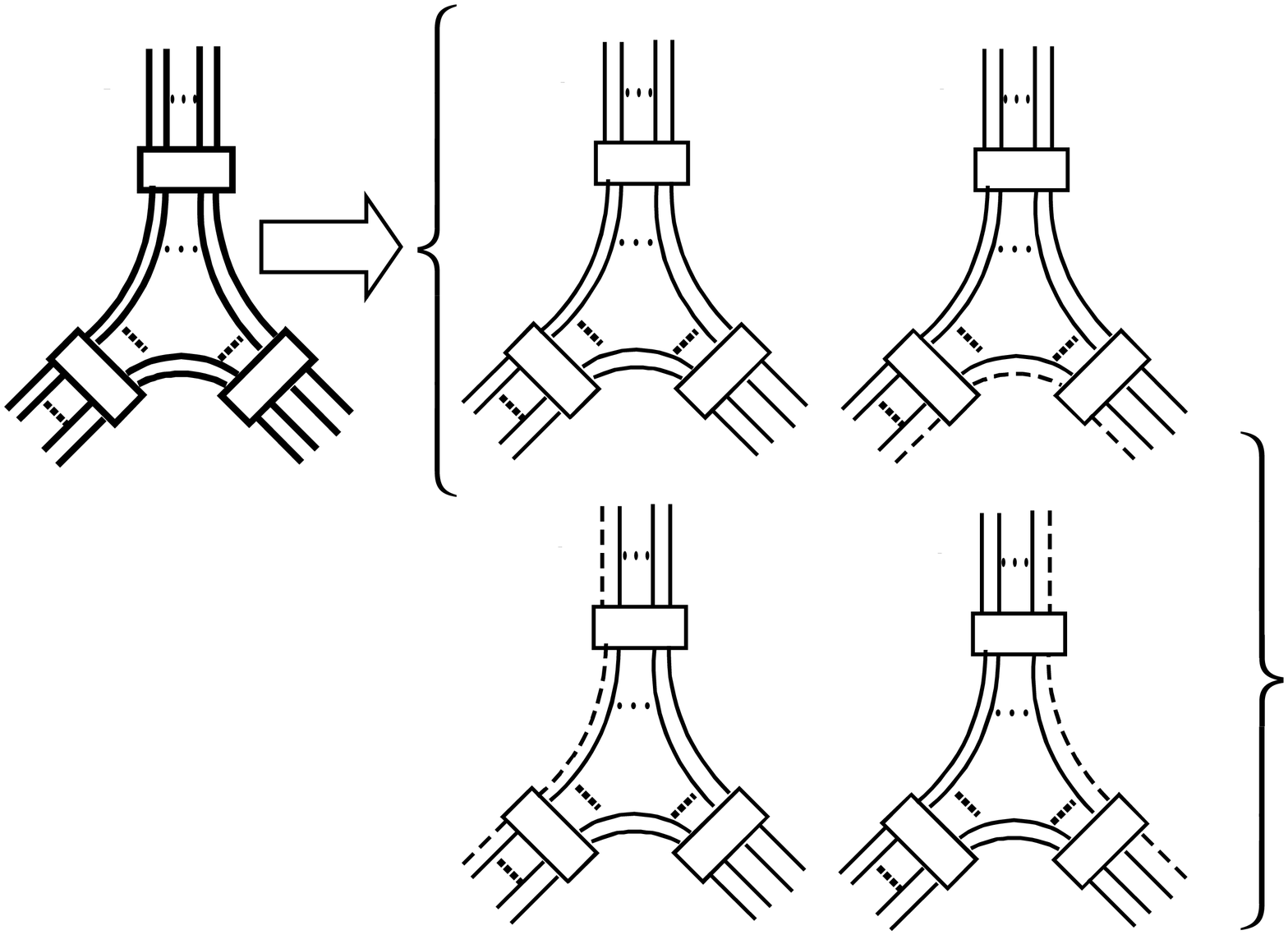}
\caption{ Decomposition of trivalent vertex in which the sum of
colors is even}
\pp

\subsection{Simple Closed Diagrams: the Super-$\Theta$ Graph}

We have found that edges and nodes of superspin networks decompose
into sums of terms, each of which consists of an ordinary spin
network, perhaps dressed by dotted lines.  As a result any closed
super spin network can be decomposed into a sum of such terms. As
an example, we describe the simplest example of a closed spin
network, which is the  $\Theta$ graph.  The simplest one is the
diagram in which every link has color one. This super  $\Theta$ graph
can be decomposed into a sum of three components, each of which is
an ordinary spin network.  This is illustrated in fig.9.

\p
\includegraphics[angle=0,width=8cm,height=4cm]{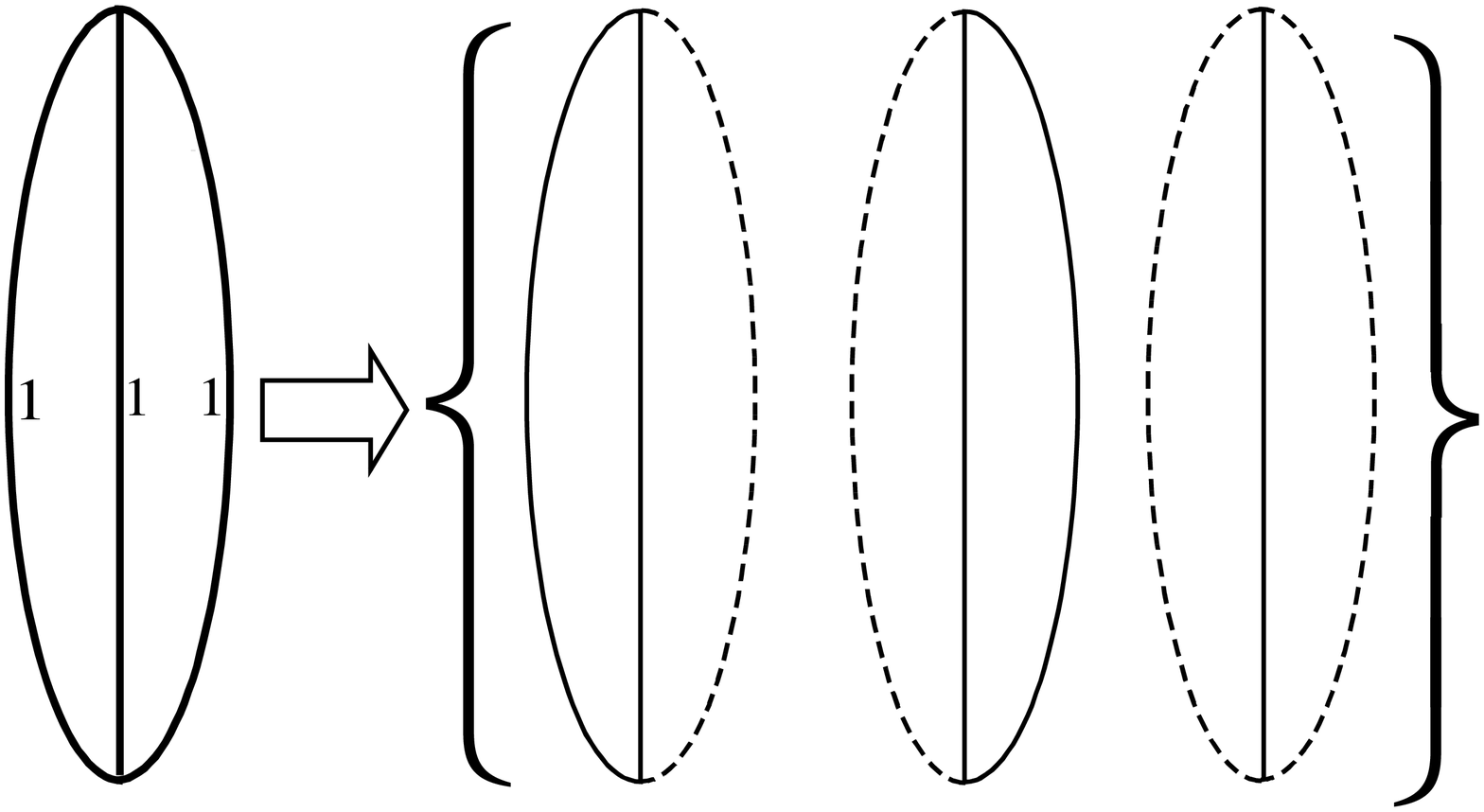}
\caption{Decomposition of the simplest  $\Theta$ graph }
\pp

Another interesting  $\Theta$ diagram is the one in which the colors
of three links are $(n,2,n)$. We will use it later in the
calculation of the area spectrum in quantum supergravity. It can
be decomposed into four components in terms of $SU(2)$ spin
networks as shown in fig.10.
\p
\includegraphics[angle=0,width=10cm,height=4cm]{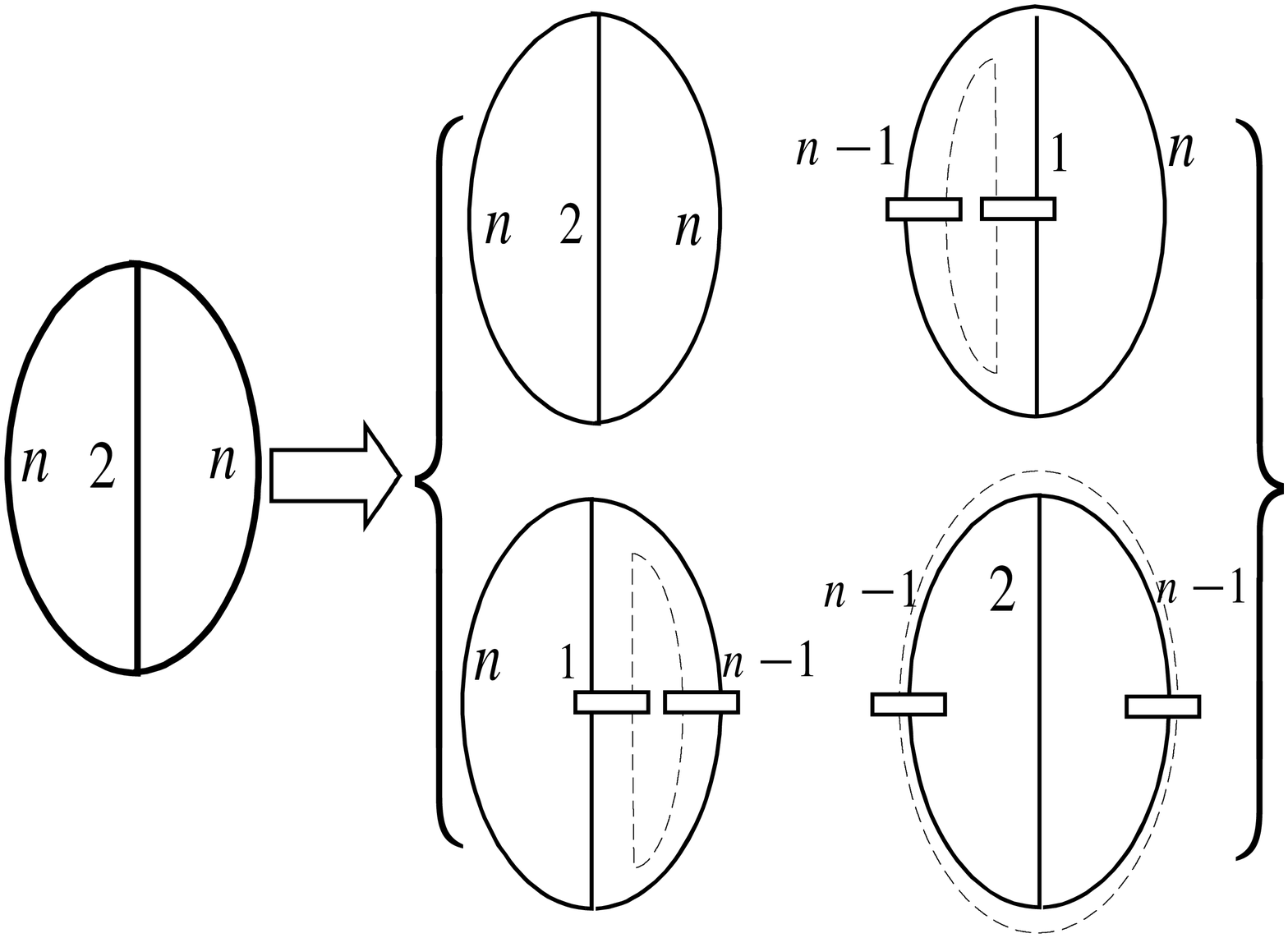}
\caption{The general case of the decomposition of  $\Theta$ graph }
\pp

\section{Evaluation of Super Spin Networks}

In the case of $SU(2)$ spin networks, the edges represent
projection operators, which live in the Temperly Lieb algebra.
These can always be decomposed using the bracket identity [see
fig.11].
\p
\includegraphics[angle=0,width=10cm,height=4cm]{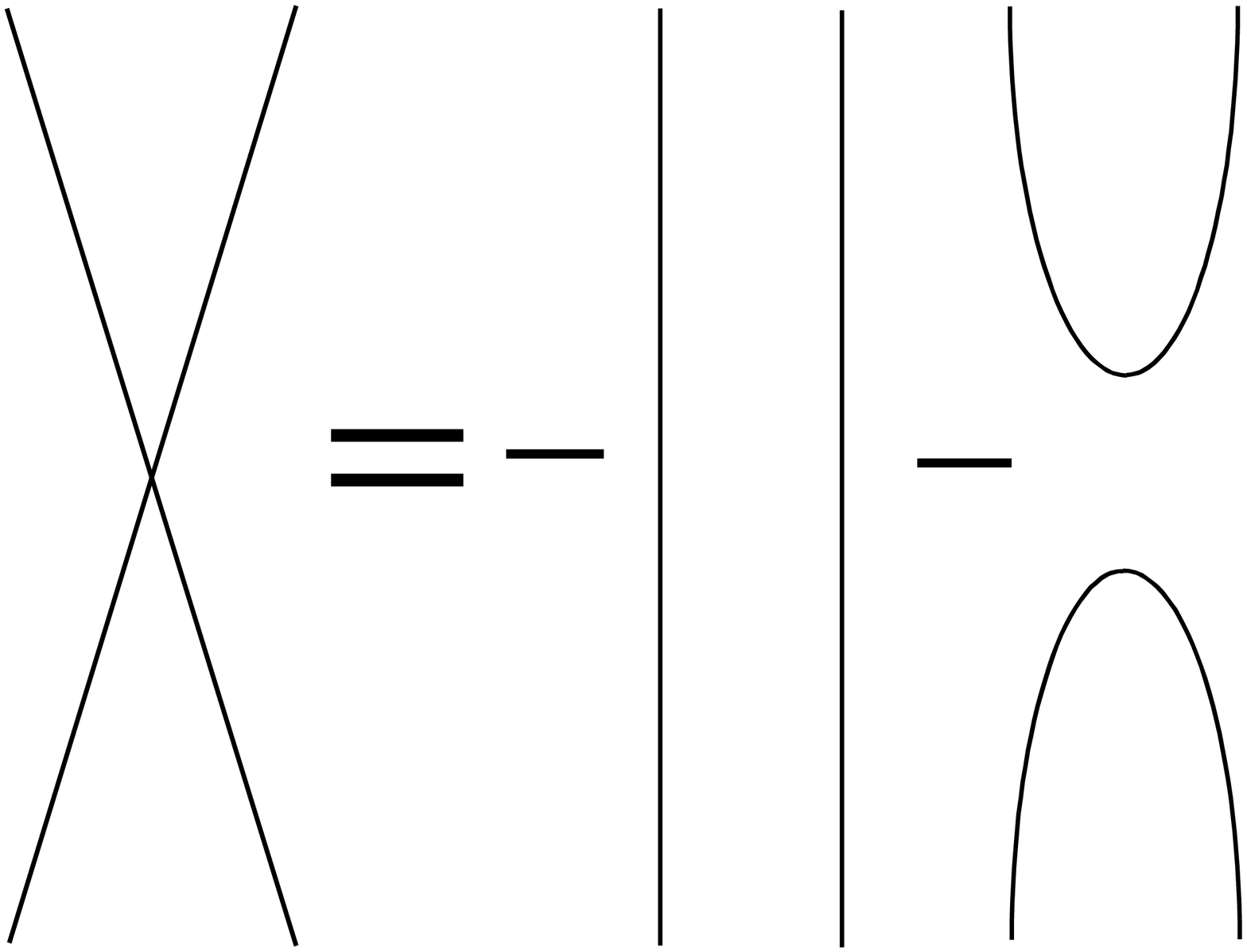}
\caption{The bracket identity for $SU(2)$ spin networks} \pp

As a result, associated with any ordinary spin network there is a number
which is called its evaluation.  This was first introduced by
Penrose\cite{pen1}. It is now known to be a special case of the
Kauffman bracket polynomial when the quantum deformation parameter
$A= \pm 1$.

For the supergroup $Osp(1|2)$, the spinor identity does not exist
any more. Therefore there is no bracket identity (although in the
super loop representation some identities analogous to the
Mandelstam identity can be expressed by means of the supertraces
of the holonomies \cite{SUG2}). But we can still evaluate a super
spin networks by first decomposing it into ordinary $SU(2)$ spin
networks, using the rules defined in the previous section, and
then evaluating each component.

The evaluation of a super spin network in fact corresponds to
taking the supertrace of a product of projection operators on the
direct product of a number of fundamental representations.  The
fact that it can be expressed in terms of the evaluations of
ordinary spin network is a consequence of the fact that the
supertrace can be decomposed into a sum of traces over the $SU(2)$
representations that make up a representation of $Osp(1|2)$.  In
fact, the sign factors necessary to turn a sum of traces into a
supertrace are already built into our formalism by the sign
factors that go into the graded symmetrizations that define the
edges and nodes of the super spin networks. In the example of the
super $\Theta$ graph, as well as in the examples that follow, one
can see how this works explicitly.

As a direct application, we can calculate the super standard
closure of the super tangles, which is defined as the supertrace
of the holonomy of the flat connection in this representation:

\begin{eqnarray}
Str_j(
\delta_{\alpha_1\alpha_2...\alpha_n}^{\alpha'_1\alpha'_2...\alpha'_n})
&=& tr_j(
\delta_{A_1A_2...A_n}^{A'_1A'_2...A'_n})+tr_j(\delta_{A_1A_2...A_{n-1}0}
^{A'_1A'_2...A'_{n-1}0})\nonumber\\
&=&(-1)^{(2j)}(2j+1)+(-1)^{2(j-1/2)}[2(j-1/2)+1]
\nonumber\\&=&(-1)^{2j} \label{Str}
\end{eqnarray}

Here when we take the trace
of the dotted line, we find its value is one.

Let us consider the simple super $\Theta$ graph in which the colors of
links are $(1,2,1)$. After decomposing the graph into the normal
spin networks and taking the trace of them as shown in fig.12, we
find the value of the  $\Theta$ graph is one.
\p
\includegraphics[angle=0,width=8cm,height=5cm]{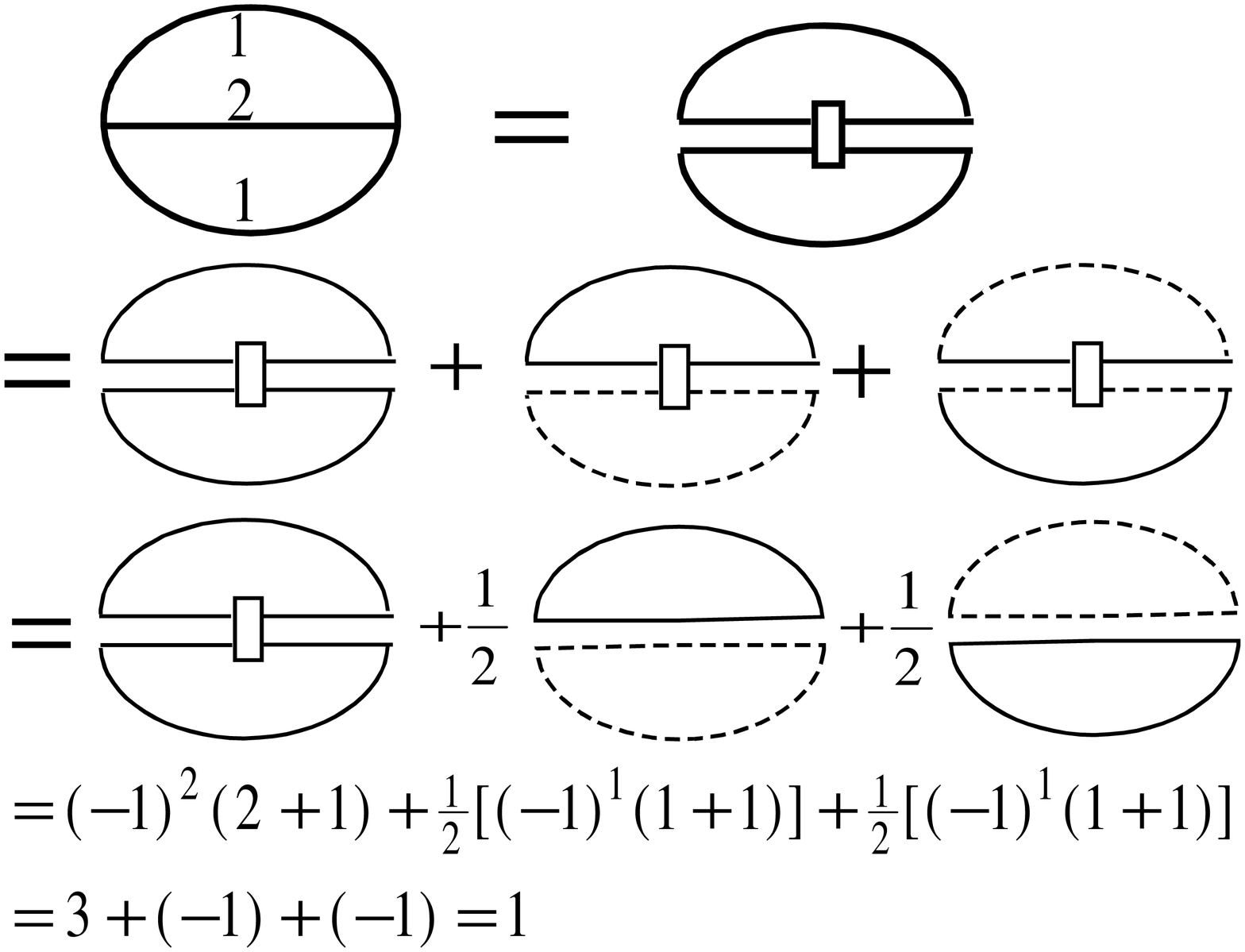}
\caption{The Evaluation of  $\Theta (1,2,1)$} \pp Also since
this graph is equivalent to the super standard closure with color
two (see the first step in figure 12), we can find the value of
this graph by (40) directly,  in which $n$ equals two. If we
consider another example in which the colors are (2,2,2), we have
the answer illustrated in fig.13.
\p
\includegraphics[angle=0,width=8cm,height=5cm]{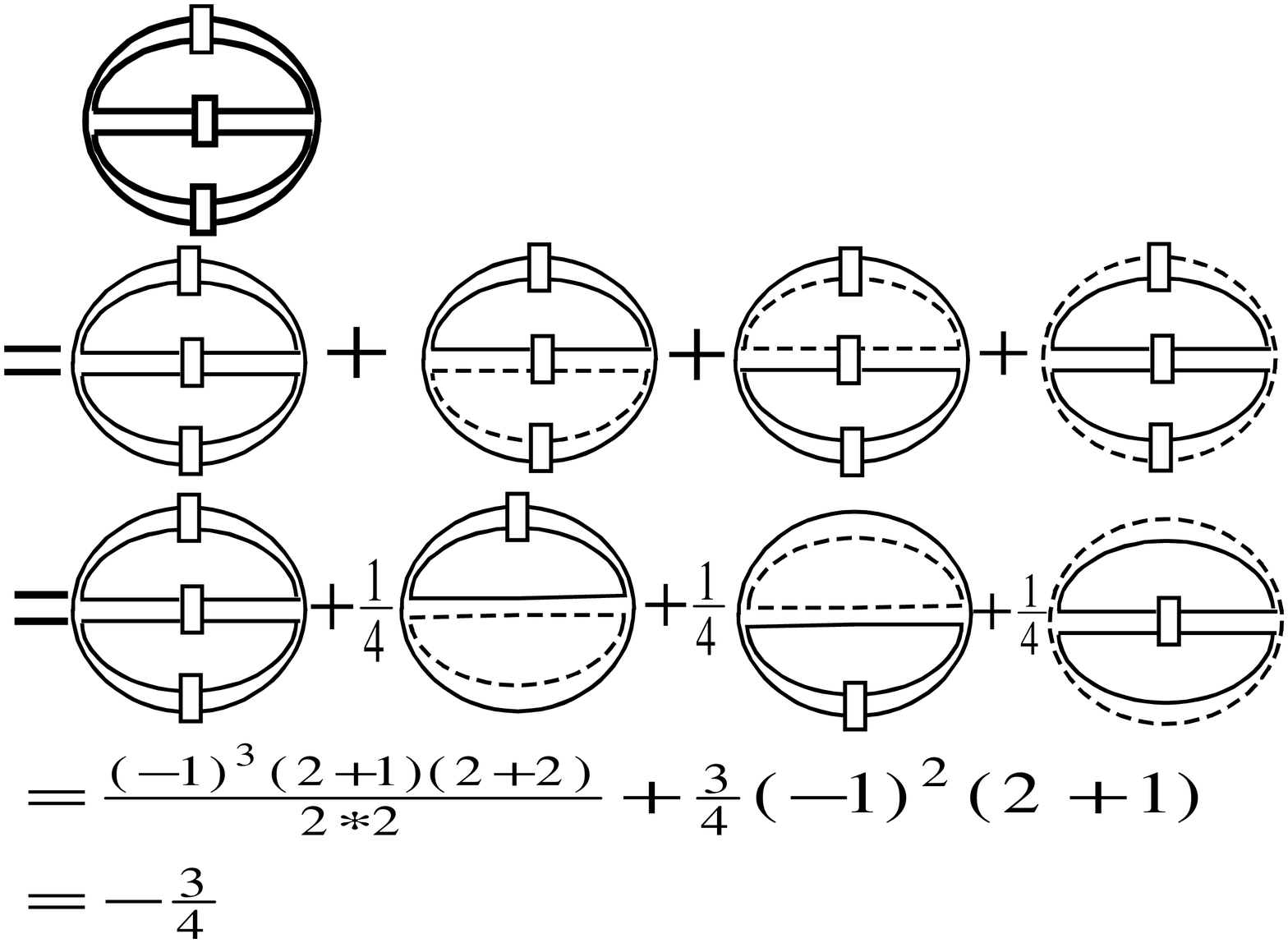}
\caption{The Evaluation of $\Theta (2,2,2)$.} \pp

In the third step the coefficient one fourth appears when we try
to separate the dotted loop from the real-line loops. Since
initially the bosonic index is symmetrized with the fermionic
indices, we have four different ways to connect the ropes to form
loops.  But the value of any loop which is formed by connecting one
real rope and one dotted rope must be zero, therefore only one
graph has non-zero value. We illustrate the specific expansion in
fig.14.
\p
\includegraphics[angle=0,width=8cm,height=5cm]{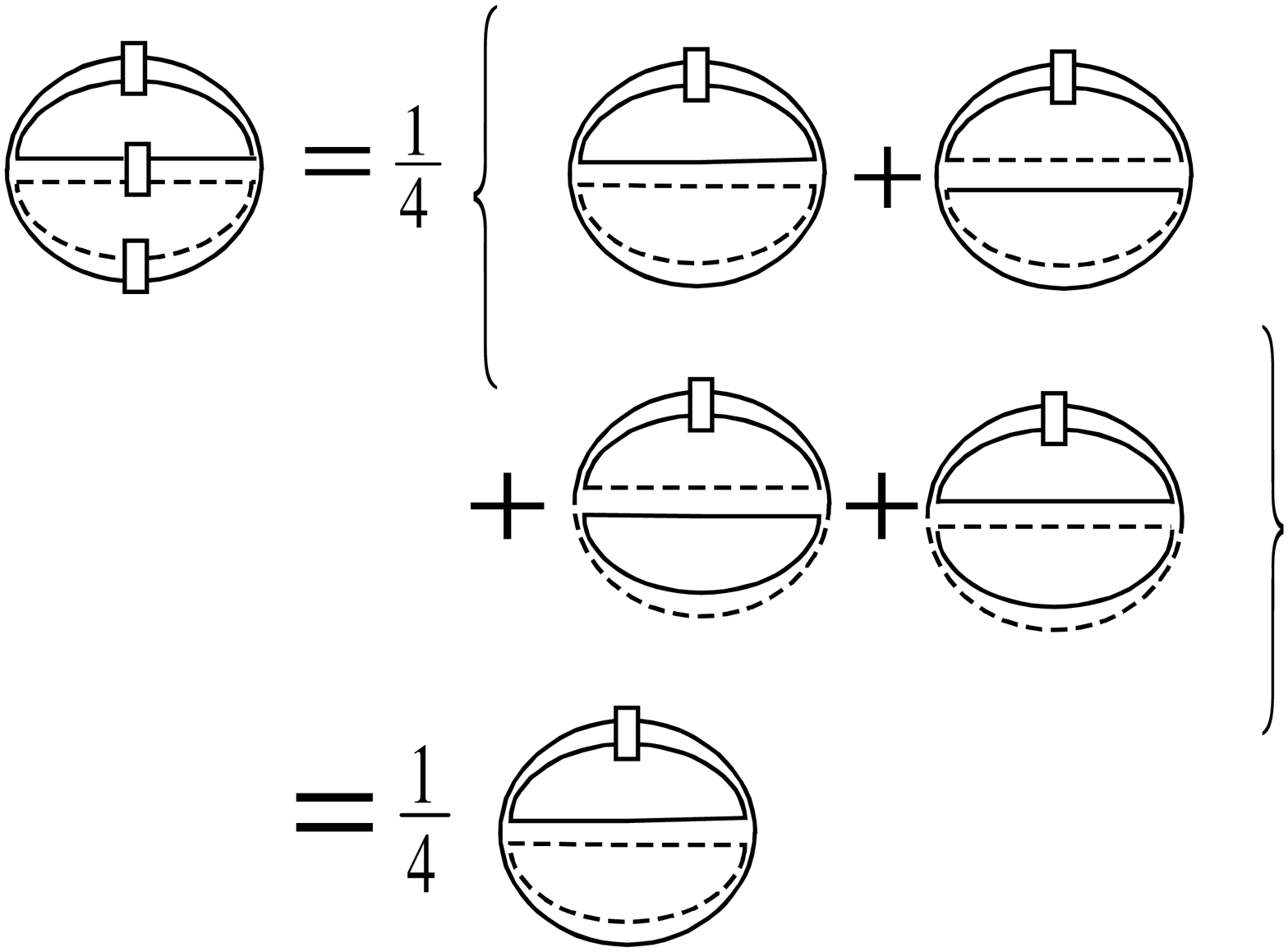}
\caption{Separation of the dotted loop from the  $\Theta$ graph}
\pp
The most interesting  $\Theta$ graph, which has important application
to the calculation of the area spectrum in supergravity, is the
one with colors (n,2,n). From the last section, we see this graph
can be divided into four graphs with respect to the ordinary
su(2) spin networks. In fig.16 the bosonic index is symmetrized
with the fermionic indices. To evaluate all these graphs, we also
need to separate the dotted loop from each  $\Theta$ graph. In
other words, we must decouple the dotted line from the
symmetrizer just as we have done for the  $\Theta$ graph
(1,2,1)and (2,2,2).
\p
\includegraphics[angle=0,width=8cm,height=5cm]{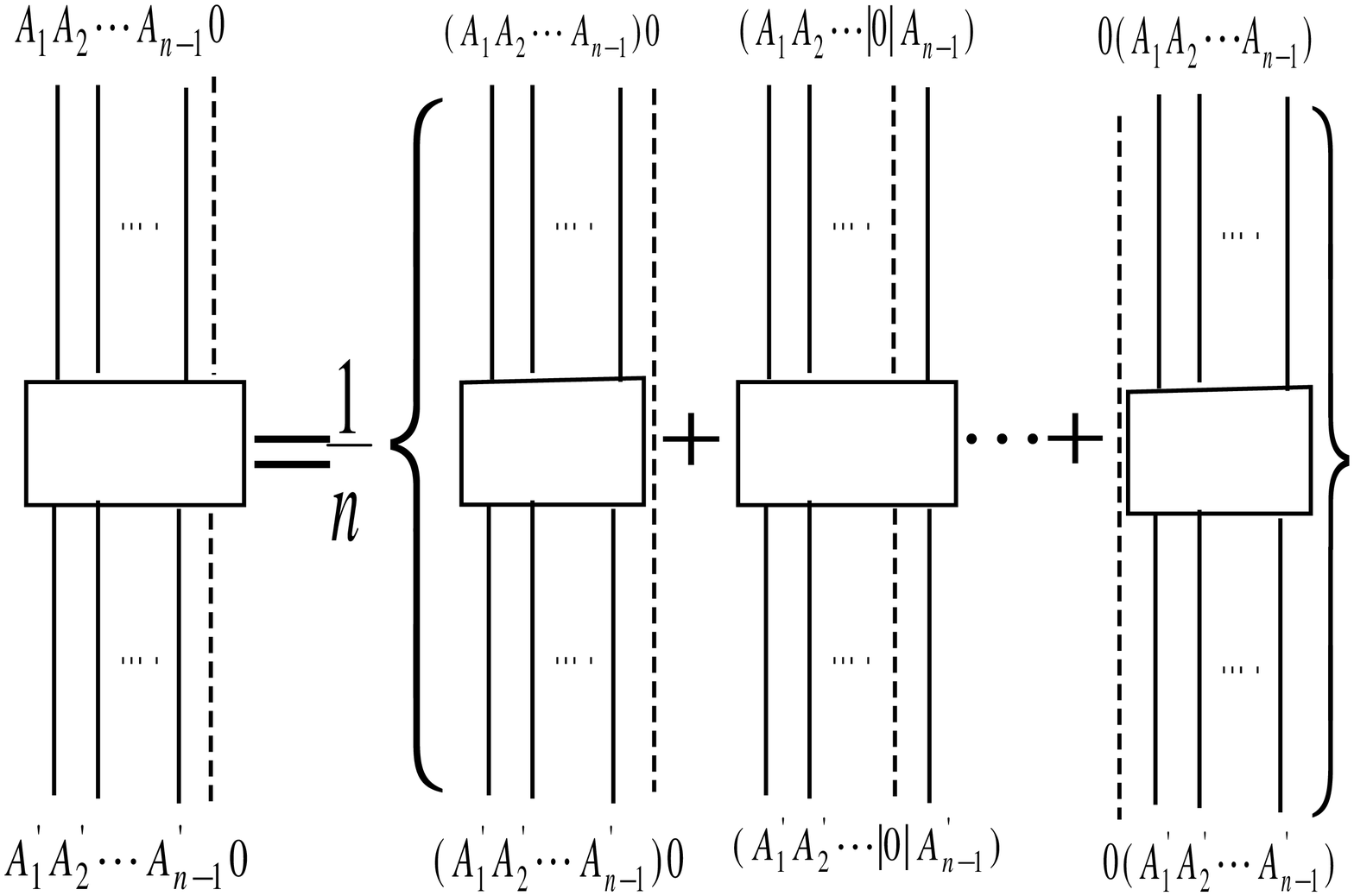}
\caption{Separation of the dotted line from the symmetrizer.}
\pp
\p
\includegraphics[angle=0,width=8cm,height=5cm]{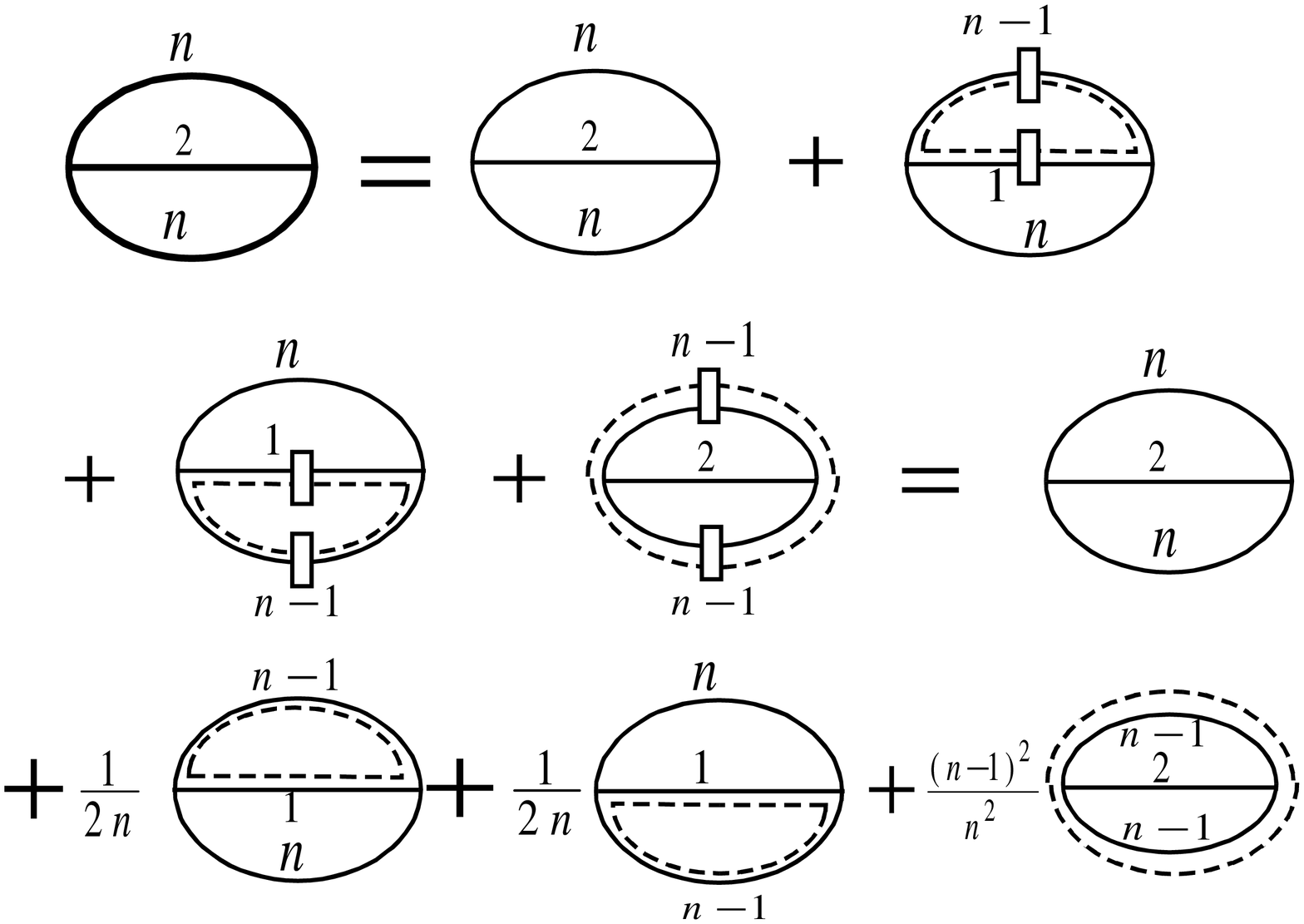}
\caption{Evaluation of the $\Theta$ graph with color (n,2,n)}
\pp

When doing this one must be careful to obtain the right
coefficients for each term. The key point is that the dotted line has to be
connected to the dotted line and for every vertex the triangle
inequality must hold. In general when the dotted line is separated
from the symmetrizer of color $n$, the factor $1/n$ appears and
there are $n$ terms  due to the different permutation of the
dotted line as shown in fig.15. For the second and third term in
figure 16, there is only one possible routing of the
dotted line which has non-zero value
among the $2n$ possibilities,  therefore the
coefficients are $1/2n$. For the last term, the dotted line can't
be connected to the link of color two so there are $(n-1)(n-1)$
routings with non-zero value  and the coefficient is
$(n-1)^2/n^2$.
Now it's straightforward to evaluate the super
 $\Theta$ graph by summing all the ordinary  $\Theta$ graphs:
\begin{eqnarray}
\Theta (n,2,n)&=&\frac{(-1)^{n+1}(n+1)(n+2)}{2n}+\frac{1}{2n}(-1)^n(n+1)
\nonumber\\&&
+\frac{1}{2n}(-1)^n(n+1)+\frac{(n-1)^2}{n^2}\frac{(-1)^nn(n+1)}{2(n-1)}
\nonumber\\& =&\frac{(-1)^{(n+1)}(n+1)}{2n} \label{eva}
\end{eqnarray}

It is not difficult to generalize this calculation to the case in which
the super $\Theta$ graph has color$(m,2,n)$. One finds that the coefficients
before every ordinary  $\Theta$ graphs respectively are
$(1,1/2n,1/2m,(n-1)(m-1)/mn)$. However, it is more
complicated to find a
general formula for the super $\Theta$ graph with color $(m,n,p)$, in
which the separation of the dotted loop from links labeled by
$m,n$ obviously depends on the third link with color $p$.

\section{The Super-area Operator and Its Spectrum}

A natural question concerning  the spin network states of
supergravity is whether we
can construct observables such as the area and the volume
of the space in terms of
their action on super spin network states, as in the case
of general relativity[4].
Here we show that the answer is yes, if the operator is
suitably defined. In this
section we construct the area operator and calculating its
eigenvalues in the context
of super spin network basis.

The gauge invariance of supergravity includes the $Osp(1|2)$
symmetry, hence we must require the observables should be
invariant under its full action. The expression for the area
operator in quantum general relativity, computed in
\cite{spain,sn1,sn2}, is not an observable in supergravity, since
it is {\it not} $Osp(1|2)$ gauge invariant. But it is not
difficult to extend the definition of the area of a surface in
general relativity to an expression which is $Osp(1|2)$ invariant.
Given a spatial surface $\cal S\mit$, which is a two-dimensional
manifold embedded in the spacetime manifold $\cal M\mit$, we define
the supersymmetric
area to be:
\f
A[\cal S\mit]=\int_{\cal s\mit}d^2s\sqrt{n_an_b\cal
E\mit^{aI}\cal E\mit^b_I}
\ff
where $n_a$ is the normal vector
of the surface and the ${\cal E}^{aI}$ is the conjugate momentum.
The definition of area operator is closely related to the two-hand
loop operator $\cal T\mit^{ab}$ that we have introduced in section three.
When the loop shrinks to a point, following \cite{vol2} and using
the identity about the supertrace of the  $Osp(1|2)$ Lie algebra
we find,
\f \cal T\mit^{ab}[\alpha](s,t)=Str[\cal
U\mit_{\alpha}(s,t) \cal E\mit^a(\alpha(t)) \cal
U\mit_{\alpha}(t,s)\cal E\mit^b(\alpha(s))]=2\cal E\mit^{aI}\cal
E\mit^b_I 
\ff
As a result, the area of the small surface with
side L, to zeroth order, can be written as,
\f
A[\cal
S\mit]=\lim_{L\rightarrow\infty}\sum_{I}\sqrt{A_I^2} \ff where: \f
A_I^2=\frac{1}{2}\int_{s_I}d^2\sigma \int_{s_I}d^2\tau
n_a(\sigma)n_b(\tau)\cal T\mit^{ab}
[\alpha_{\sigma\tau}](\sigma,\tau)
\ff

Now we define the $Osp(1|2)$ invariant area operator to be,
\f
\hat{A}[{\cal S}]=\lim_{L\rightarrow\infty}\sum_{I}
\sqrt{\hat{A}_I^2}
\ff
where,
\f
\hat{A}_I^2=\frac{1}{2}\int_{s_I}d^2\sigma\int_{s_I} d^2\tau
n_a(\sigma)n_b(\tau) \hat{\cal T}^{ab}
[\alpha_{\sigma\tau}](\sigma,\tau)
\ff

Next we want to consider
the action of the area operator on spin network states. In
\cite{sn1,vol2}, the discrete spectrum of the area operator in
spin network states is worked out in different ways. One can
divide the link, the element of the spin networks, into ropes in
loop representation so that the area operator acts on the state as
a second order loop operator which can be expressed in terms of
the elementary grasp operation; or equivalently one can define the
action of the area operator on spin networks as inserting two
trivalent intersections on the link by a new link of color $2$,
then calculate the eigenvalues of the operator by recoupling
theory directly. Here we can define the action of super $\cal T\mit$
variables in terms of the elementary grasp operation.  This
allows us to calculate the spectrum of the operator
in both ways.

Let us consider the former method first. The action of the super
operator $\cal T\mit^a$ on the super spin networks can be defined
as fig.(17).
\p
\includegraphics[angle=0,width=8cm,height=5cm]{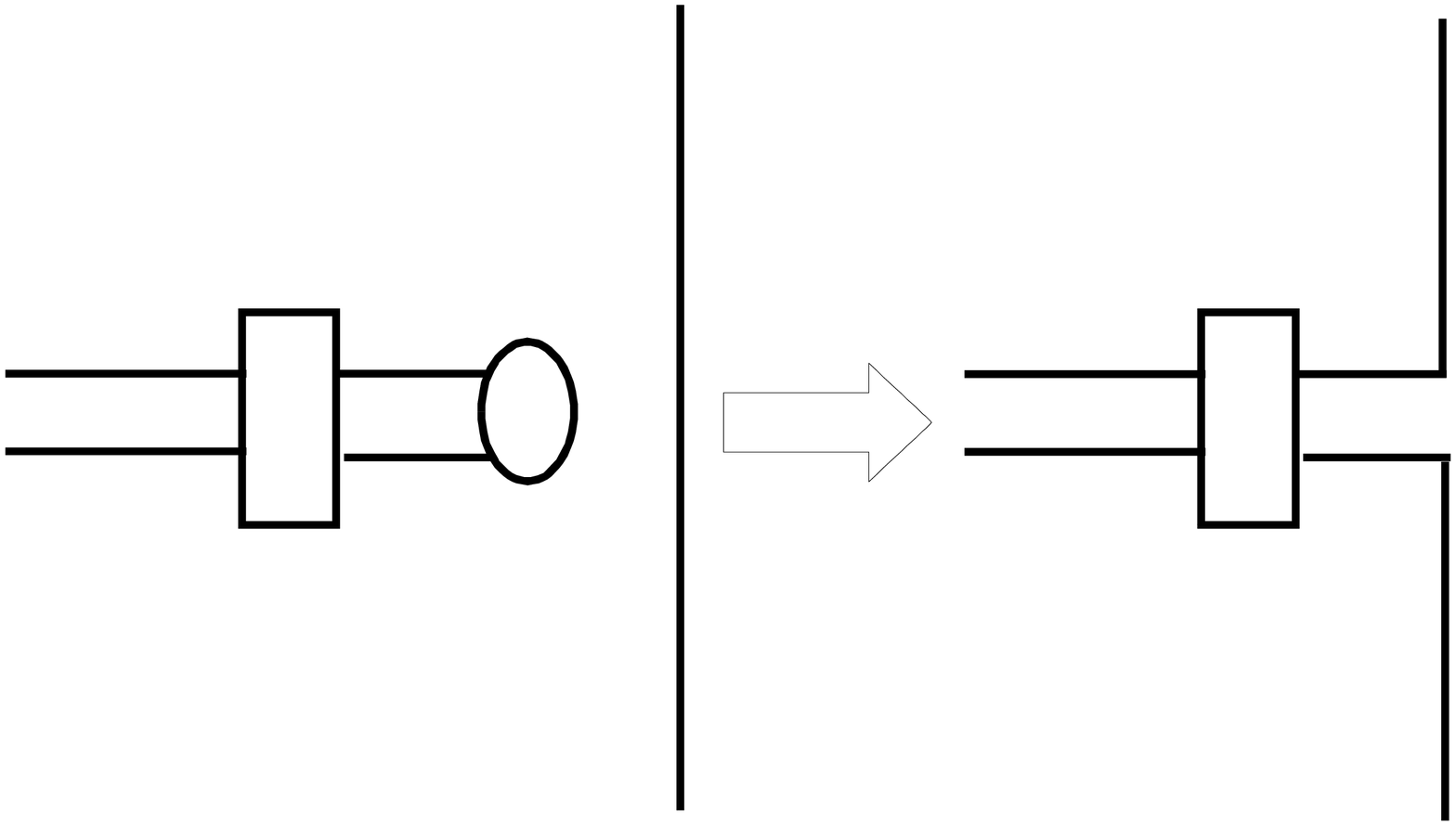}
\caption{Action of $\cal T\mit^a$ with one grasp on the super spin
networks }
\pp
Basically as we have done in the previous
sections, we can decompose super spin networks into the ordinary
SU(2) ones and then consider the action of the operator on them
separately. From fig.4, we see the super link of color 2 can be
divided into two components, so the corresponding action of the
super operator can be divided into two parts which can be
illustrated in figure 18. For convenience,  let's define these
actions as ``real grasp'' and ``dotted grasp'' respectively.

\p
\includegraphics[angle=0,width=8cm,height=5cm]{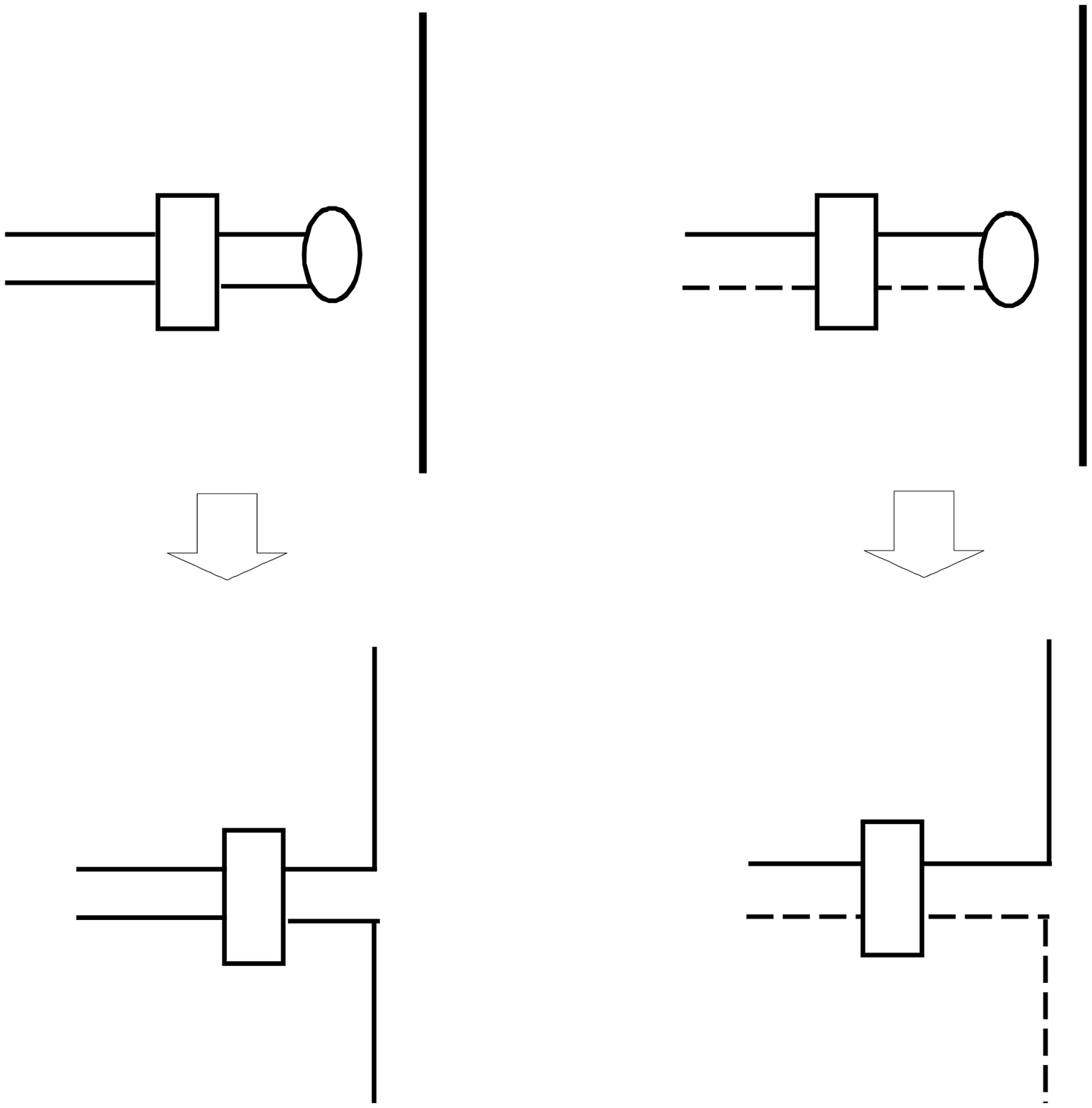}
\caption{The super grasps in the view of ordinary SU(2) spin
networks}
\pp

When decomposed in terms of $SU(2)$ spin networks, we find that
there are several distinct grasp operations. The first possibility
is the real grasp to the real line, which is exactly the normal
grasps having appeared in \cite{vol2}. The second one is the
dotted grasp acting on the real line,  and the third one the
dotted grasp acting on the dotted line. Note that the real grasp
acting on the dotted line vanishes since the only possible result
is that two real lines combine together and go back.

Now it's straightforward to express the action of super operators
on the link of color n, but we need be careful to determine the
multiplicative factors when using the Leibnitz rule to define the
action of area operator on it. Specially, there is a great
difference between the real grasp and the dotted grasp. Since the
area operator is related to the second order super loop operator,
we can take the $\cal T\mit^{ab}$ as the handle with two grasps in
the super spin network basis. When two grasps act on the link of
color $n$, they can grasp the same rope, or any two different
ropes,  so there are $n^2$ possible ways to grasp the link. But
when the real grasps act on the dotted rope, the results of the
action are zero. So the number of ``non-zero'' grasps are $n^2$ and
$(n-1)^2$ to the doublet of the super link respectively. Also
after the two dotted grasps act on the link of color $n$, we need
separate the dotted rope from the solid ropes so that we can apply
the formula with respect to the ordinary $SU(2)$ spin networks. As
we have discussed in the last section, the separation involves the
factor $1/2n$. As a result the coefficients before the graphs
acted by the dotted grasps are $n/2$. Figure 19 and 20 show the
actions of these two kinds of grasps on the link of color n.
\p
\includegraphics[angle=0,width=8cm,height=5cm]{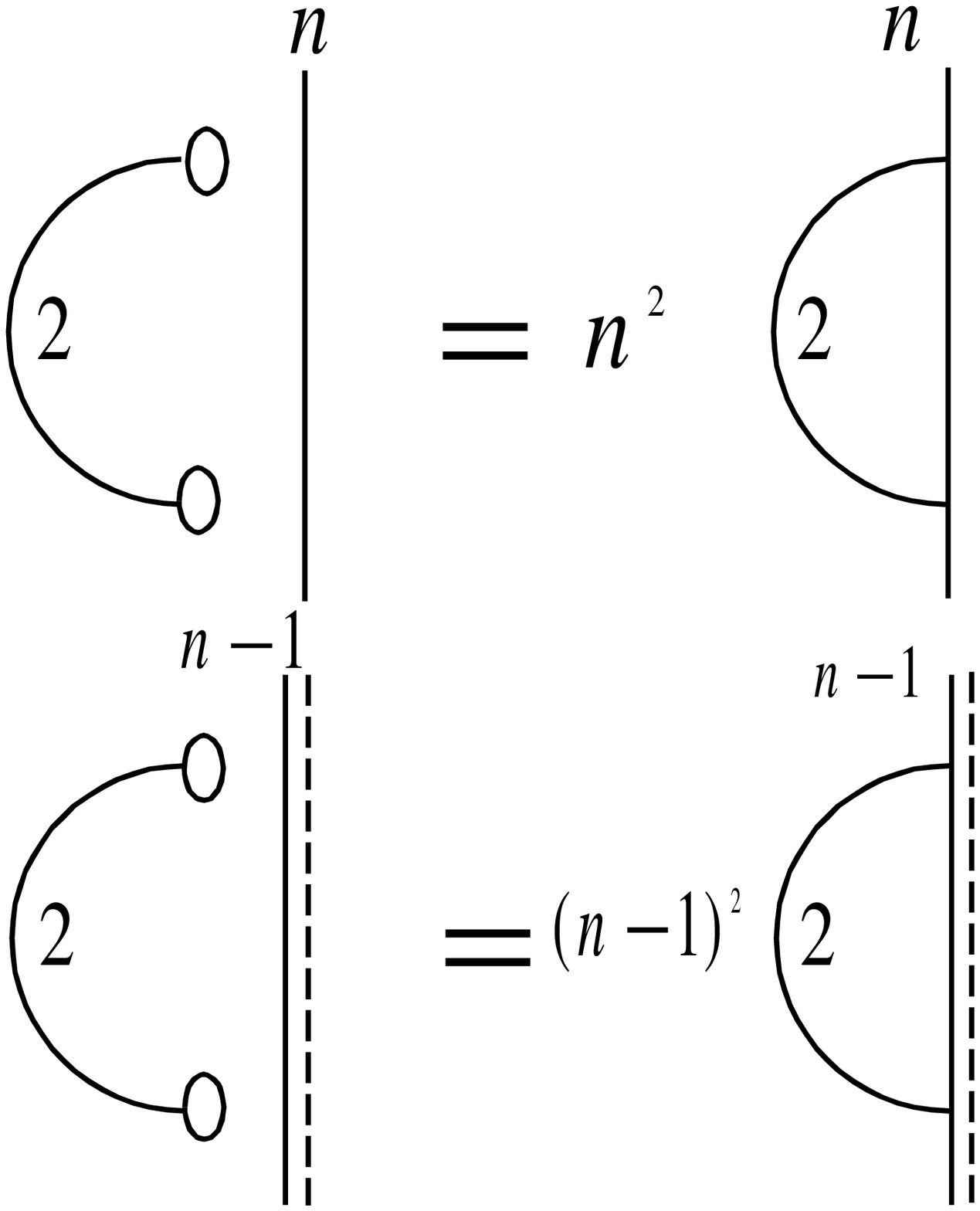}
\caption{Action of the second order real grasp}
\pp
\p
\includegraphics[angle=0,width=8cm,height=5cm]{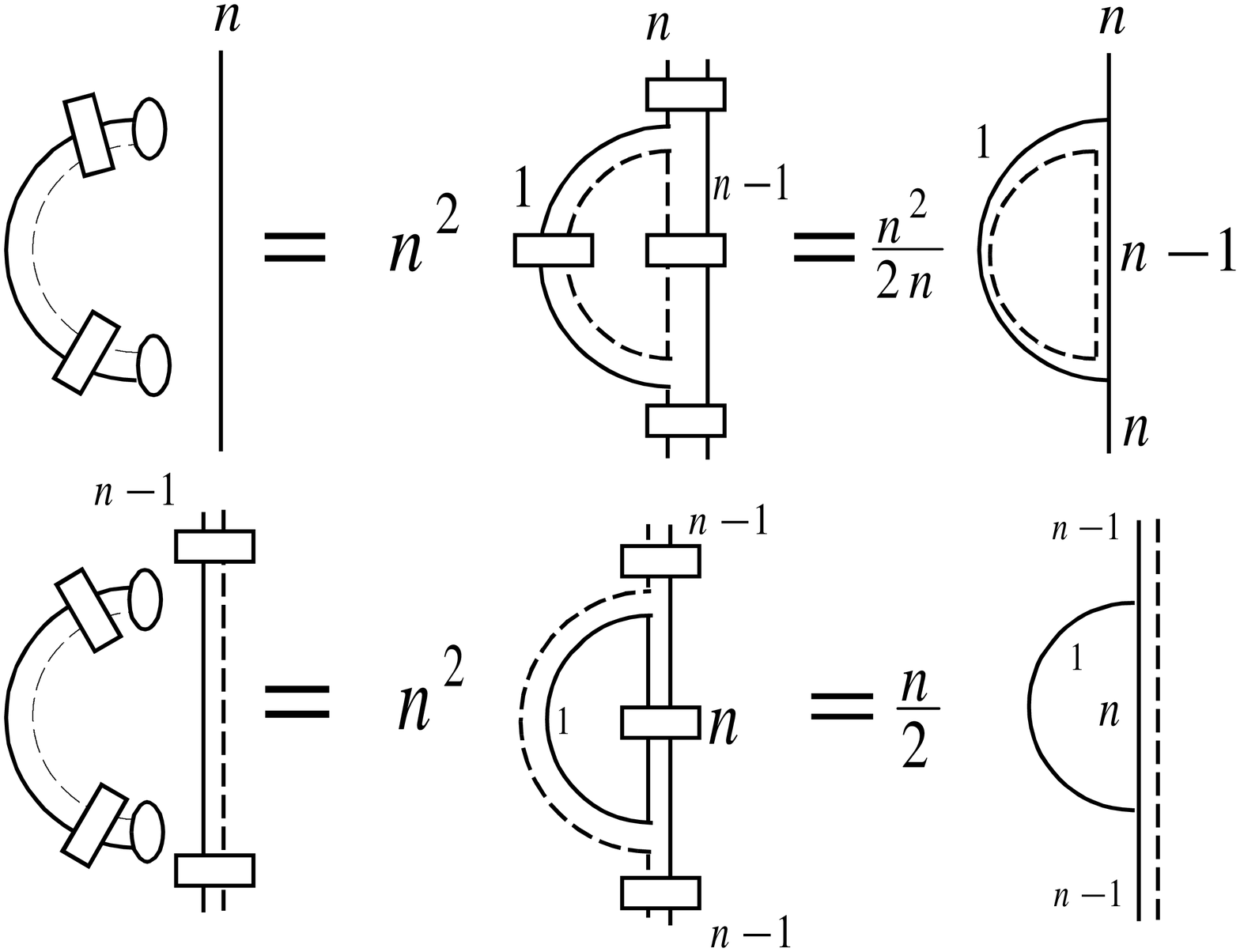}
\caption{Action of the second order dotted grasp }
\pp

Finally we arrive at the last step of this section, that is to
calculate  the spectrum of the area operator. Combining the two
actions of the grasps together, we find the super spin network
states are the eigenstates of the $\hat{A}^2$. It is
straightforward to compute the eigenvalues of $\cal T\mit^{ab}$ in
the super spin network basis (see fig.21) and the result is:
\p
\includegraphics[angle=0,width=10cm,height=5cm]{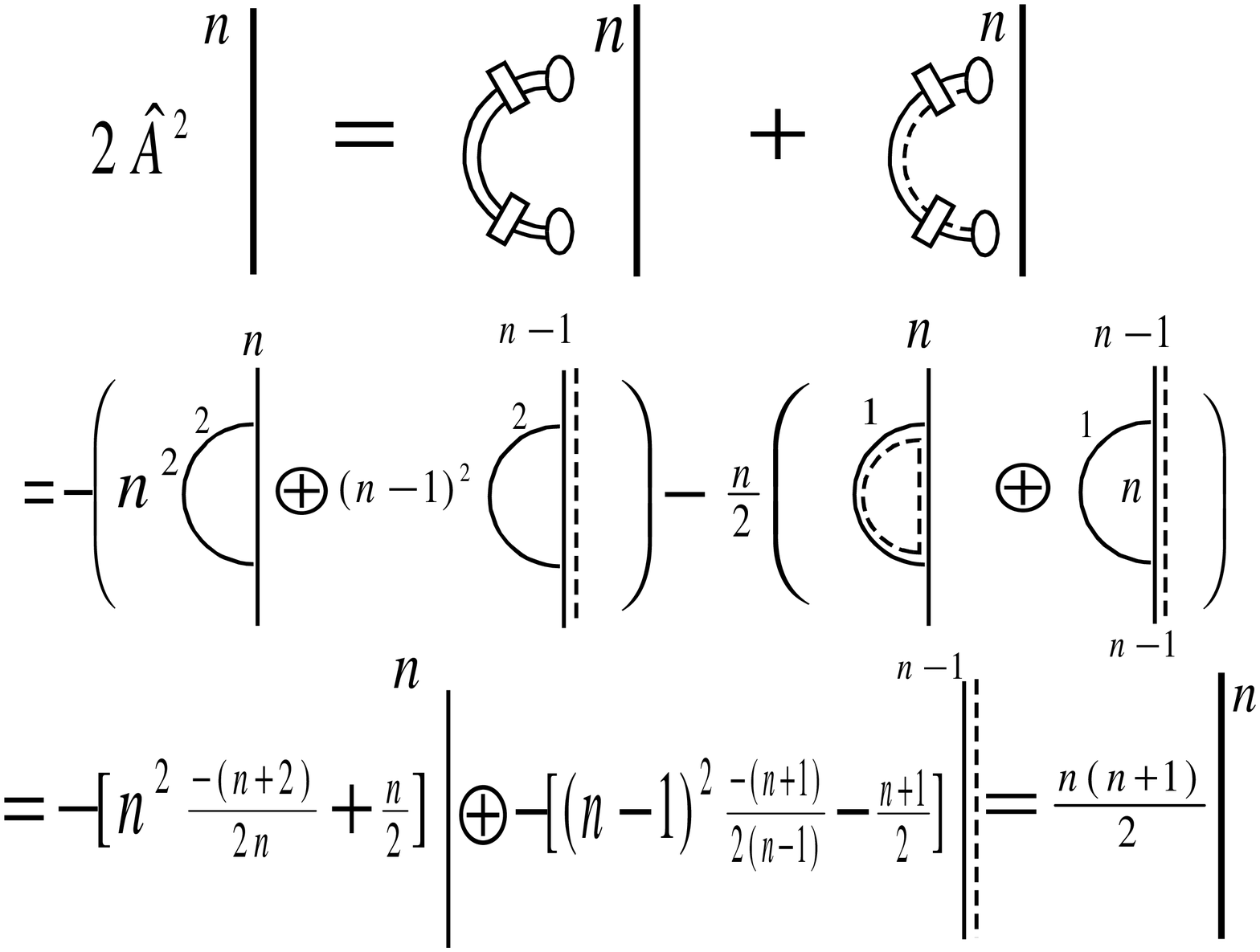}
\caption{The action of the area operator } \pp
\f \hat
{A}^2|\Gamma^{sg},n_i,v_e>=
\sum_i\frac{n_i(n_i+1)}{4}l_p^4|\Gamma^{sg},n_i,v_e> \ff

Where $l_p$ is Planck length. As a result we find that the eigenvalues of the area operator are
given by, \f \hat{A}|\Gamma^{sg},n_i,v_e>=
\sum_i\sqrt{\frac{n_i(n_i+1)}{4}}l_p^2|\Gamma^{sg},n_i,v_e>
=\sum_i\sqrt{j_i(j_i+\frac{1}{2})}l_p^2|\Gamma^{sg},n_i,v_e> \ff

Here we have applied the identities and the formulas in SU(2) spin
networks. This confirms the expected result that the spectrum is
discrete and is directly related to the eigenvalues of the Casmier
operator of $Osp(1|2)$.

Next we conclude that we can get the same solution to eigenvalues
of  the area operator directly by employing the identity
associated with the representation of $Osp(1|2)$, in which the
evaluation of the super projectors eq.(\ref{Str}) and super
$\Theta$ graphs (\ref{eva}) is involved. The procedure is shown in
fig.22.
\p
\includegraphics[angle=0,width=8cm,height=5cm]{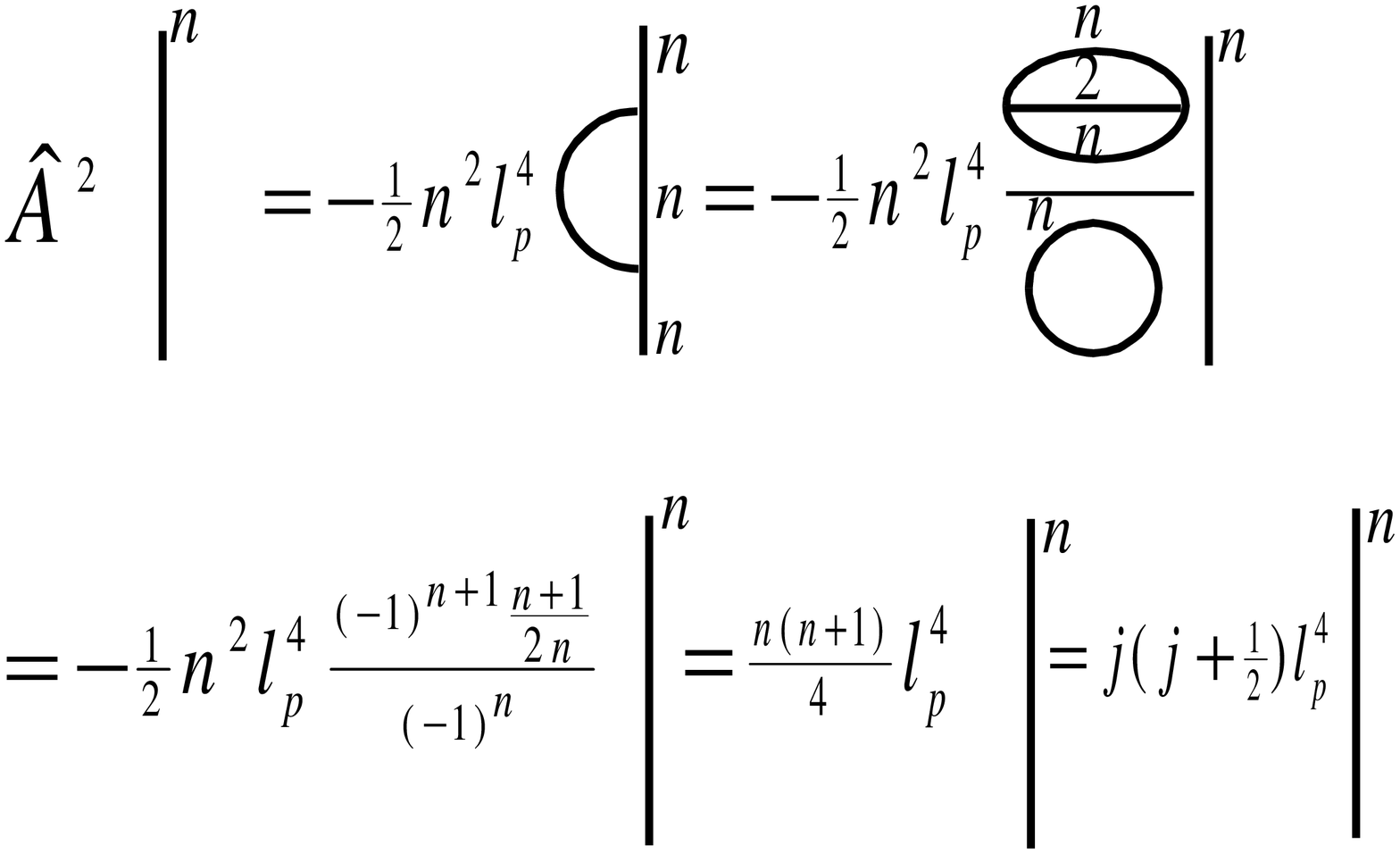}
\caption{The evaluation of area spectrum by means of the identity
for super spin networks}
\pp

Finally, we note that we have computed here only a part of the
spectrum of the superarea operator, which is that concerned with
the intersections of edges of the superspin network with the
surface $S$. As in the $SU(2)$ case there are additional
eigenvalues associated with the possibility that the surface $S$
intersects nodes of the superspin network. These eigenvalues may
not be physically relevant as the probability of such
intersections is zero, but in any case they can be computed
directly.

\section{Conclusions}

In this paper we have taken an important step in the extension of
the results of loop quantum gravity to supergravity and string
theory.  We have shown that for ${\cal N}=1$ supergravity in $3+1$
dimensions there is a straightforward extension of the methods
developed in \cite{sn1,sn2,vol2} from quantum general relativity.
The extension to ${\cal N}=2$ is in progress and will be reported
shortly \cite{yi2,SUG1}. There is in fact nothing to prevent the
direct extension to any $\cal N$, what is difficult is only the
question of whether, for ${\cal N} >2 $, all the degrees of
freedom of higher $\cal N$ supergravities are represented by an
extension of the connection representation, or whether additional
degrees of freedom need to be introduced. In this connection we
may note that the extension of the loop representation to
represent the states of $p$-form gauge fields is straightforward,
and has been worked out for $p=2$ \cite{rodolofo-p,lee-p} and
$p=3$ \cite{M11}. In the latter case an extension of the loop
representation that describes a limit of $\cal M$ theory in which
only the $3$-form field of $11$ dimensional supergravity survives
can be discussed, and a large set of exact non-perturbative states
found \cite{M11}.

Finally, the quantum deformation of the supersymmetric spin
network states may be developed along the lines of \cite{KL}, and
applied both to yield the supersymmetric extension of the spin
foam models\cite{barrettcrane1,mike} as well as to the background
independent formulation of $\cal M$ theory described in
\cite{tubes,mtheory}.

\section*{Acknowledgements}

We are grateful to Shyamoli Chaudhuri,  Laurent Freidel,
Murat Gunyadin,
Renata Loll, Fotini
Markopoulou,
Adrian Ocneanu and Mike Reisenberger
for conversations and encouragement.
This work was supported by the NSF through grant
PHY95-14240
and a gift from the Jesse Phillips Foundation.


\begin{thebibliography}{99}

\bibitem{KS}J. Kogut and L. Susskind, {\it Hamiltonian formulation of 
Wilson's lattice gauge theories}, Phys. Rev. D11: 395,1975.

\bibitem{spain}L. Smolin, in {\it Quantum Gravity and
Cosmology}, eds  J  P\'erez-Mercader {\it et al}, World Scientific,
Singapore 1992;  {\it The future of spin networks},
 gr-qc/9702030 in the Penrose Feshscrift.

\bibitem{sn1}C. Rovelli and L. Smolin,
{\it Discreteness of area and volume in quantum gravity},
 Nuclear Physics B 442 (1995) 593.  Erratum: Nucl. Phys.
B 456 (1995) 734.

\bibitem{sn2}C. Rovelli and L. Smolin,
{\it Spin networks and quantum gravity},
 Physical Review D 52 (1995) 5743-5759, gr-qc/9505006.

\bibitem{sen}A. Sen, {\it Gravity as a spin system}, 
Phys. Lett. B119 (1982) 89; {\it Quantum theory of a spin
3/2 system in Einstein spaces}, Int. J. of Theor. Phys.
21 (1982) 1.

\bibitem{abhay}A. Ashtekar, Phys. Rev. Lett.
57 (1986) 2244; Phys. Rev. D36 (1987) 1587.

\bibitem{barrettcrane1}J. Barrett and L. Crane, {\it
Relativistic spin networks and quantum gravity}, 
J. Math. Phys.39(1998) 3296-3302, gr-qc/9709028.

\bibitem{hologr}L. Smolin, {\it Holographic formulation of
quantum general relativity}, hep-th/9808191.

\bibitem{carlo-review}C. Rovelli, {\it Loop Quantum Gravity}, Review paper
written for the electronic journal `Living Reviews', gr-qc/9710008.

\bibitem{future}L. Smolin, {\it The
future of spin networks}, in the Penrose Feshscrift, gr-qc/9702030.

\bibitem{vol2}R. Loll, Nucl. Phys. B444 (1995) 619;
B460 (1996) 143; R. DePietri and C. Rovelli, {\it Geometry
eigenvalues and
scalar product from recoupling theory in loop quantum gravity},
gr-qc/9602023, Phys. Rev. D54 (1996) 2664; 
Simonetta Frittelli, Luis Lehner, Carlo Rovelli,
{\it The complete spectrum of the area from recoupling theory
in loop quantum gravity},
gr-qc/9608043; R. Borissov, Ph.D. thesis, Temple, (1996).

\bibitem{superstuff}T. Jaocobson, {\it New variables for canonical
supergravity} Class. Quant. Grav. 5 (1988) 923.

\bibitem{GSU1}
T. Sano, J. Shiraishi, {\em The Non-perturbative Canonical Quantization of the
N=1 Supergravity}, Nucl. Phys. B410(1993) 423, hep-th/9211104; H. Kunitomo, 
T. Sano, {\em The Ashtekar formulation for
Canonical N=2 Supergravity} Int. J. Mod. Phys.D1(1993)559; {\em The Ashtekar formalism and WKB Wave Functions of
N=1,2 Supergravities}, hep-th/9211103.
 H. Kunitomo and T. Sano
{\it The Ashtekar formulation
for canonical N=2 supergravity}, Prog. Theor. Phys. suppl. (1993) 31; 
T. Sano and J. Shiraishi, {\it The Non-perturbative Canonical
Quantization of the N=1 Supergravity}, Nucl. Phys. B410 (1993) 423,
hep-th/9211104; {\it The Ashtekar Formalism and WKB Wave Functions of
N=1,2 Supergravities}, hep-th/9211103; K. Ezawa, {\it  Ashtekar's
formulation for N=1,2 supergravities as ``constrained'' BF theories
}, Prog. Theor. Phys.95(1996) 863-882, hep-th/9511047.

\bibitem{GSU2}
K. Ezawa, {\em Ashtekar's formulation for N=1,2 supergravities as
``constrained'' BF theories}, Prog. Theor. Phys.95(1996) 863-882, 
hep-th/9511047.

\bibitem{SUG2}
D. Armand-Ugon, R. Gambini, O. Obregon, J. Pullin, {\em Towards a loop
representation for quantum canonical supergravity}, Nucl. Phys. B460(1996) 
615-631, gr-qc/9508036; 
R. Graham, C. Csordas, {\em Exact quantum state for N=1
supergravity}, Phys. Rev. D52(1995) 6656-6659, gr-qc/9507008.

\bibitem{N=4}T. Kadoyoshi
and S. Nojiri,
{\it N=3 and N=4 two form supergravities}, Mod. Phys. Lett. A12: 
1165-1174,1997,
hep-th/9703149; L. F.
Urrutia {\it Towards a loop representation of connection theories
defined over a super-lie algebra}, hep-th/9609001.

\bibitem{pen1}
R.Penrose, {\em The theory of quantized directions}, in Quantum
theory and beyond, ed T.Bastin, Cambridge U Press 1971.

\bibitem{bae1}
J. Baez, {\em Spin Networks in Nonperturbative Quantum
Gravity}, in The Interface of Knots and Physics, ed. Louis Kauffman,
A.M.S. Providence,1996, 167-203, gr-qc/9504036; {\em Spin Network States in Gauge theory}, Adv. Math. 117(1996)253-272, gr-qc/9411007.

\bibitem{string-bh}
J. Maldacena, {\it Black Holes in String Theory}, hep-th/9607235; 
A. W. Peet, {\it The Bekenstein Formula and String Theory},
 Class. Quant. Grav.15(1998) 3291-3338, hep-th/9712253; G. T. Horowitz and J. Polchiski, {\it Correspondence
Principle for Black Holes and Strings}, Phys. Rev. D55, 6189(1997).

\bibitem{kirill-bh}K. Krasnov, gr-qc/9603025,
Phys.Rev. D55 (1997) 3505-3513; 
gr-qc/9605047, Gen. Rel. Grav. 30 (1998) 53-68;
A. Ashtekar, J. Baez, A. Corichi, K. Krasnov,
Phys. Rev. Lett. 80 (1998) 904-907, gr-qc/9710007;
 K. Krasnov, Class. Quant. Grav.16(1999)L15-L18, gr-qc/9902015.

\bibitem{giovanni}
G. Amelino-Camelia, {\it Could we observe the desciteness of
quantum gravity length and area operators?}, gr-qc/9808047.

\bibitem{yi2}Y. Ling, L. Smolin, {\it Holography, BPS states, and
N=2 supergravity}, In preparation.

\bibitem{yi3}Y. Ling, In preparation.

\bibitem{SUG1}
Y. Ling, L. Smolin, {\em Holographic Formulation of Supergravity},
In preparation.

\bibitem{linking}L. Smolin, {\it Linking topological quantum field
theory and nonperturbative quantum gravity}
gr-qc/9505028, J. Math. Phys. 36 (1995) 6417.

\bibitem{rayner}D. Rayner, Class. Quan. Grav. 7 (1990)111; 7 (1990) 651.

\bibitem{chrisabhay}A. Ashtekar and C. J. Isham,
 Class. Quant. Grav. 9 (1992) 1069.

\bibitem{gangof5}A. Ashtekar, J. Lewandowski, D. Marlof, J. 
Mour\~{a}u, T. Thiemann, {\it Quantization of diffeomorphism
invariant theories of connections with local degrees of
freedom}, gr-qc/9504018, J. Math. Phys. 36 (1995) 519; 
A. Ashtekar and J. Lewandowski, {\it Quantum
Geometry I: area operator}, Class. Quant. Grav. 14(1997)A55-A82, gr-qc/9602046;
J. Lewandowski, {\it Volume and quantization}
, Class. Quant. Grav. 14(1997)71-76, gr-qc/9602035.

\bibitem{thomas}T. Thiemann, {\it Quantum Spin Dynamics I-VI},
Class. Quant. Grav. 15 (1998) 839-905, 1207-1314, 1463-1485,
gr-qc/9606092, gr-qc/9606089, gr-qc/9606090,
gr-qc/9705020, gr-qc/9705019, gr-qc/9705018,
gr-qc/9705017; {\it Kinematical Hilbert Spaces for Fermionic and Higgs Quantum
Field Theories}, Class. Quant. Grav. 15(1998)1487-1512, gr-qc/9705021.

\bibitem{GSU3}
A. Pais, V. Rittenberg, {\em Semisimple Graded Lie Algebras},
J. Math. Phys.16 (1975)2062; Err.ibid.17 (1976)598.

\bibitem{GSU4}
P. Minnaert, M. Mozrzymas {\em Algebra Structure of tensor
superoperators for the super-rotation algebra I and II}, 
J. Math. Phys. 33(1992)1582, 1594.

\bibitem{GSU5}
A. B. Balantekin, I. Bars, {\em Dimension and character formulas for
Lie supergroups}, J. Math. Phys. 22(1981)1149; {\em Representation of
supergroups}, J. Math. Phys. 22 (1981)1810.

\bibitem{GSU6}
L. F. Urrutia, H. Waelbroeck, F. Zertuche, {\em The Algebra of
Supertraces for (2+1) Super De Sitter Gravity}, 
Mod. Phys. Lett. A7(1992)2715-2721.

\bibitem{GSU7}
P. Minnaert, S. Toshev, {\em Racah Sum Rule and Biedenharn-Elliott
Identity for the Super-Rotational 6-j Symbols}, J. Math. Phys. 35(1994) 
5057-5064, hep-th/9402040.

\bibitem{KL}
L. Kauffman, S. L. Lins, {\em Temperley-Lieb Recoupling Theory and
Invariants of 3-Manifolds}, Princeton U Press, 1994.

\bibitem{rodolofo-p}P. Arias, C.Di Bartolo, X. Fustero, R. Gambini, 
A. Trias, {\it Quantum Abelian surfaces}, UAB-FT-148, Apr. 1986; 
{\it Second quantization of the antisymmetric potential in the Abelian 
surfaces space}, Int. J. Mod. Phys. A7:737-754,1992.

\bibitem{lee-p}L. Smolin, {\it Finite diffeomorphism invariant observables 
for quantum gravity}, Phys. Rev. D49(1994)4028, gr-qc/9302011.

\bibitem{M11}L. Smolin, {\it Chern-Simons theory in 11 dimensions as a 
non-perturbative phase of M theory}, hep-th/9703174.

\bibitem{mike}M. Reisenberger,
{\it A lattice worldsheet sum for 4-d Euclidean general
relativity}, gr-qc/9711052; M. Reisenberger and C. Rovelli,
{\it ``Sum over Surfaces'' form of Loop Quantum Gravity},
 Phys. Rev. D56(1997)3490-3508, gr-qc/9612035.

\bibitem{tubes}F. Markopoulou and L. Smolin, 
{\it Quantum geometry with intrinsic local causality}
preprint, Dec. 1997, gr-qc/9712067, Phys. Rev. D58 (1998) 084032.

\bibitem{mtheory}L. Smolin, {\it A candidate for a
background independent formulation of $\cal  M$ theory}, hep-th/9903166.

\end{thebibliography}
\end{document}